\documentclass{aa}  

\usepackage{graphicx}
\usepackage{lscape}
\usepackage{longtable}
\usepackage[table]{xcolor}    
\usepackage{txfonts}
\usepackage[breaklinks=true]{hyperref}
\hypersetup{
    bookmarksopen=false,
    bookmarksnumbered=false,
    unicode=true,           
    pdftoolbar=true,        
    pdfmenubar=true,        
    pdffitwindow=false,     
    pdfstartview={FitH},    
    pdfnewwindow=true,      
    colorlinks=true,       
    linkcolor=red,          
    citecolor=magenta,        
    filecolor=green,      
    urlcolor=blue           
}

\usepackage{amssymb}
\usepackage{pifont}
\usepackage{todonotes}
\newcounter{todocounter}

\usepackage{amsmath}
\usepackage{multirow}

\begin{document}

   \title{Testing the chemical tagging technique with open clusters
        \thanks{Based on observations obtained at the Telescope Bernard Lyot (USR5026) operated by the Observatoire Midi-Pyrénées, Université de Toulouse (Paul Sabatier), Centre National de la Recherche Scientifique of France, and on public data obtained from the ESO Science Archive Facility under requests  number 81252 and 81618.}
   }

   \author{ S. Blanco-Cuaresma \inst{1, 2}
            \and C. Soubiran \inst{1}
            \and U. Heiter \inst{3} 
            \and M. Asplund \inst{4}
            \and G. Carraro \inst{5}
            \and M. T. Costado \inst{6}
            \and S. Feltzing \inst{7}
            \and J. I. Gonz\'alez-Hern\'andez \inst{8, 9, 10}
            \and F. Jim\'enez-Esteban \inst{11, 12}
            \and A. J. Korn \inst{3}
            \and A. F. Marino \inst{4}
            \and D. Montes \inst{8}
            \and I. San Roman \inst{13}
            \and H. M. Tabernero \inst{8}
            \and G. Tautvai\v{s}ien\.{e} \inst{14}
          }
   \offprints{S. Blanco-Cuaresma, \email{Sergi.Blanco@unige.ch}}

   \institute{CNRS / Univ. Bordeaux, LAB, UMR 5804, F-33270, Floirac, France.
        \and
            Observatoire de Gen\`eve, Universit\'e de Gen\`eve, CH-1290 Versoix, Switzerland
        \and 
            Department of Physics and Astronomy,  Uppsala University, Box 516, 75120 Uppsala, Sweden
        \and
            Research School of Astronomy and Astrophysics, Australian National University, ACT 2601, Australia
        \and 
            ESO, Alonso de Cordova 3107, Casilla 19001, Santiago de Chile, Chile
        \and 
            Instituto de Astrof\'isica de Andaluc\'ia-CSIC, Apdo. 3004, 18080, Granada, Spain
        \and
            Lund Observatory, Department of Astronomy and Theoretical Physics, Box 43, SE-221 00 Lund, Sweden
        \and
            Dpto. Astrof\'isica, Facultad de CC. F\'isicas, Universidad Complutense de Madrid, E-28040 Madrid, Spain.
        \and 
            Instituto de Astrof\'isica de Canarias, E-38205 La Laguna, Tenerife, Spain.
        \and 
            Universidad de La Laguna, Dept. Astrof\'isica, E-38206 La Laguna, Tenerife, Spain.
        \and
            Centro de Astrobiolog\'{\i}a (INTA-CSIC), Dpto. de Astrof\'{\i}sica, PO Box 78, E-28691, Villanueva de la Ca\~nada, Madrid, Spain 
        \and 
            Suffolk University, Madrid Campus, C/ Valle de la Viña 3, 28003, Madrid, Spain 
        \and
            Centro de Estudios de F\'isica del Cosmos de Aragon, Plaza San Juan 1, Planta-2, 44001, Teruel, Spain
        \and
            Institute of Theoretical Physics and Astronomy, Vilnius University, Gostauto 12, 01108 Vilnius, Lithuania
   }


 
  \abstract
    {Stars are born together from giant molecular clouds and, if we assume that the priors were chemically homogeneous and well-mixed, we expect them to share the same chemical composition. Most of the stellar aggregates are disrupted while orbiting the Galaxy and most of the dynamic information is lost, thus the only possibility of reconstructing the stellar formation history is to analyze the chemical abundances that we observe today.}
   {The chemical tagging technique aims to recover disrupted stellar clusters based merely on their chemical composition. We evaluate the viability of this technique to recover co-natal stars that are no longer gravitationally bound.}
   {Open clusters are co-natal aggregates that have managed to survive together. We compiled stellar spectra from 31 old and intermediate-age open clusters, homogeneously derived atmospheric parameters, and 17 abundance species, and applied machine learning algorithms to group the stars based on their chemical composition. This approach allows us to evaluate the viability and efficiency of the chemical tagging technique.}
   {We found that stars at different evolutionary stages have distinct chemical patterns that may be due to NLTE effects, atomic diffusion, mixing, and biases. When separating stars into dwarfs and giants, we observed that a few open clusters show distinct chemical signatures while the majority show a high degree of overlap. This limits the recovery of co-natal aggregates by applying the chemical tagging technique. Nevertheless, there is room for improvement if more elements are included and models are improved.
   }
   {}


   \maketitle
%


\section{Introduction}

Understanding the formation and evolution of galaxies and their structure (e.g., disks) is an open issue in near-field cosmology. One approach to tackle this problem is to study our own Galaxy by unravelling the sequence of events that took place in the formation of the Galactic disk (where most star formation occurs). Unfortunately, most of the dynamical information is lost since the disk was formed in a dissipative process and it evolved dynamically. Nevertheless, the chemical composition of the stars can potentially help us to recover the history of our Galaxy \citep{2002ARA&A..40..487F}.

Stars are born in aggregates from molecular clouds \citep{1987ARA&A..25...23S, 2000prpl.conf..121M, 2003ARA&A..41...57L}. Hydrodynamical simulations indicate that the progenitor cloud undergoes fragmentation preventing contraction onto a single star \citep{2005A&A...435..611J, 2004MNRAS.353..769T, 1995MNRAS.272..213L}. Hundreds to thousands of stars can be formed from one single cloud. If we assume that the progenitor cloud was chemically well-mixed \citep{2014arXiv1408.6543F}, then we expect to observe homogeneous chemical composition in the stars formed from this cloud \citep{2010ApJ...713..166B}. With this information, we could use the method of chemical tagging to track individual stars back to their common formation sites as proposed by \cite{2002ARA&A..40..487F}.

The viability of this approach depends on two critical issues: Do stars born together have the same chemical signature? And, are the chemical signatures different enough to distinguish stars formed from different molecular clouds?

In this context, open clusters are a fantastic laboratory to investigate their homogeneity and validate that their progenitor clouds were uniformly mixed. Most of the stars formed in clusters in our Galaxy have already been dispersed into the field, but a few of them managed to stay gravitationally bound, probably, thanks to a higher formation rate and/or galactic orbits that avoided high-density regions and giant molecular clouds \citep{1995ARA&A..33..381F, 1995AAS...18710702F, 1980A&A....88..360V}. Thus, old and intermediate-age clusters (age >$\sim$100 Myr) are the leftovers of star forming aggregates in the Galactic disk that have managed to survive until the present day. We can be certain that their stars were born from the same molecular cloud at roughly the same period.

Additionally, since each molecular cloud has its own history of pollution by ejecta from core-collapse supernovae (i.e., Type II, Ib, and Ic Supernovae where most of the $\alpha$-elements are produced), Type Ia supernovae (SNe Ia where most iron peak elements are created), and asymptotic giant branch stars (AGB where a s-process takes place), we expect different open clusters to have different chemical patterns.

There is some observational evidence for chemical homogeneity in open clusters. For instance, \cite{2006AJ....131..455D} present an abundance analysis of the heavy elements Zr, Ba, La, Ce, and Nd (their abundances are not thought to be modified during normal stellar evolution) for F-K dwarfs in the Hyades open cluster. They claimed that the abundances of member stars are highly uniform and they showed a scatter on the order of 0.06 dex for Zr; 0.05 dex for Ba; 0.03 dex for Ce, La, and Nd. 

In a subsequent study, \cite{2007AJ....133.1161D} measured the lighter elements Na, Mg, Si, Ca, Mn, Fe, and Ni (but not for Zr and Ba) for 12 red giants of the old open cluster Collinder 261. They demonstrated again a high chemical homogeneity for this cluster finding a dispersion of 0.07 dex for Na, 0.05 dex for Mg and Ca, 0.06 dex for Si, 0.03 dex for Mn, 0.02 dex for Fe, and 0.04 dex for Ni. Additionally, they compared Collinder 261, the Hyades, and the HR 1614 moving group and were able to show that the three have unique chemical signatures. 

This last study was extended in \cite{2009PASA...26...11D}, where the authors compiled abundances of 24 open clusters from the literature and showed that different clusters seem to have different chemical patterns (from the average values). They observe significant dispersion for some elements; however, one possible reason is systematic uncertainties among the different studies (e.g., use of different methods, atomic data, model atmospheres).

\cite{2013MNRAS.428.2321M} quantified the level to which chemical tagging can distinguish between co-natal stars (stars born at the same period and formation site) by developing a metric and deriving an empirical probability function based on chemical abundances for 35 clusters collected from the literature. The authors showed that achieving a high clustering detection efficiency is difficult and that depends on the level of uniqueness of the co-natal stars' chemical signatures.

Although abundances are now available in the literature for many stars in open clusters, it is not appropiate to mix them into a single dataset to study the chemical patterns in the open cluster population (as it was done in most of the previous studies). The abundances have been obtained by diverse observers using different instruments and methods, resulting in possible systematic differences. The use of different methods (e.g., equivalent width or synthetic spectral fitting), atomic data, model atmospheres, and continuum normalization processes can lead to systematic errors. Other studies derive very accurate chemical abundances but in a non-automatic way (e.g., manual continuum and/or line fitting). Manual analysis is affected by subjective criteria that can vary with time and is not easily used when a huge quantity of spectra must be analyzed. For instance, the on-going Gaia-ESO Public Spectroscopic Survey \citep[GES;][]{2012Msngr.147...25G} will target approximately 100,000 field and open cluster stars in the Galaxy, and the GALactic Archaeology \citep[GALAH; ][]{2015arXiv150204767D} with HERMES \citep{2010SPIE.7735E...8B} survey will target a million disk stars at high resolution in a relatively short time-span. 

It is worth noting that we do not intend to build an exhaustive compilation of past and on-going studies in this introduction. We note, however, some other examples of significant contributions to the determination of chemical abundances for large samples of open clusters: the Bologna Open Clusters Chemical Evolution project (BOCCE, \citealt{2006AJ....131.1544B}); the WIYN Open Cluster Study \citep{2000ASPC..198..517M, 2011AJ....142...59J}; and recent papers from GES such as \cite{2014A&A...563A..44M}, where the chemical homogeneity of the inner-disk open clusters Trumpler 20, NGC4815, and NGC6705 from the first GES data was shown.

To evaluate the potential of chemical tagging when using our own homogeneous and automatic analysis, we have

\begin{enumerate}
    \item collected high-resolution spectra of open clusters' stars observed by different instruments and homogenized them;
    \item implemented a completely automatic process to derive atmospheric parameters and chemical abundances;
    \item applied machine learning algorithms to try to recover the original clusters from the homogeneously derived chemical abundances.
\end{enumerate}

We describe the collected data in Sect.~\ref{s:observations}. The spectral analysis developed to derive the atmospheric parameters and chemical abundances is presented in Sect.~\ref{s:spectral_analysis}. In Sect.~\ref{s:chemical_tagging} we explore the results of our analysis to validate the viability of the chemical tagging technique and, finally, the conclusions can be found in Sect.~\ref{s:conclusions}.

\section{Sample selection and observations}
\label{s:observations}

We compiled 2,133 high-resolution spectra of which 146 come from the NARVAL instrument, 1,630 from HARPS, and 357 from UVES. The initial selection criteria were that 
the spectral resolution had to be at least 47,000 (to match the setup of the GES) and the star had to be located in the field of view of a cluster (i.e., inside a given radius around the cluster center). We mainly looked for clusters discussed in \cite{2010A&A...517A..32P} and \cite{2014A&A...561A..93H}, although we did not strictly limit the selection to these. Based on this dataset, a further selection process was performed according to cluster membership and spectrum quality as described in Sect. \ref{s:data_homogenization}.

\subsection{NARVAL spectra}

The NARVAL spectropolarimeter is mounted on the 2m Telescope Bernard Lyot \citep{2003EAS.....9..105A} located at Pic du Midi (France). The data from NARVAL were reduced with the Libre-ESpRIT pipeline \citep{donati97}. These spectra were taken within a large program proposed as part of the ``Ground-based observations for Gaia" (P.I: C. Soubiran).

NARVAL spectra cover a large wavelength range ($\sim 300 - 1100$~nm), with a resolution \footnote{In this text, the term "resolution" refers to $R = \frac{\lambda}{\Delta\lambda}$ where $\lambda$ is the wavelength.} that varies for different observation dates and along the wavelength range, typically from $75,000$ around 400~nm to $85,000$ around 800~nm. However, it is acceptable to initially assume a constant resolution of R$\simeq$81,000 as we showed in \cite{2014A&A...566A..98B}.

\subsection{HARPS spectra}

HARPS is the ESO facility for the measurement of radial velocities with very high accuracy. It is fiber-fed from the Cassegrain focus of the 3.6m telescope in La Silla \citep{2003Msngr.114...20M}. The spectra were originally reduced by the HARPS Data Reduction Software (version 3.1). The data used in this work were taken from the public HARPS archive by selecting observed stars in the clusters' field of view. 

The spectral range covered is $378 - 691$~nm, but as the detector consists of a mosaic of two CCDs, one spectral order (from 530~nm to 533~nm) is lost in the gap between the two chips.

\subsection{UVES spectra}

The UVES spectrograph is hosted by unit telescope 2 of ESO's VLT \citep{2000SPIE.4008..534D}. We took the spectra available from the Advanced Data Products collection of the ESO Science Archive Facility\footnote{\url{http://archive.eso.org/eso/eso\_archive\_adp.html}}  (made available in October 2013) by selecting observed stars in clusters' field of view. 

The setup used for each observation (CD\#3, centered around 580~nm) provides a spectrum with two different parts which approximately cover the ranges from 476 to 580~nm (lower part) and from 582 to 683~nm (upper part).

\subsection{Data homogenization}
\label{s:data_homogenization}

The wavelength range varies from one set of observations to another. We chose to limit the spectral analysis to the range between 480 and 680~nm, where all the spectra provide their best signal-to-noise ratio (S/N). 

To increase the overall S/N, we co-added spectra corresponding to the same star when they were observed by the same instrument and with the same setup (i.e., same resolution). After co-addition, we discarded spectra with S/N lower than 40, which is an optimal level for determining atmospheric parameters with iSpec \citep{2014A&A...569A.111B}.

All the spectra were convolved to 47,000, which is the minimum resolution from our initial selection of spectra.

Observations were cross-correlated with a zero point template corresponding to a solar spectrum observed by NARVAL \citep{2014A&A...566A..98B}. The derived radial velocities were used to shift and align all the spectra. 

We assume that cluster members share the same velocity vector with a small random dispersion, thus we discarded stars with a radial velocity higher or lower than the cluster's reference velocity $\pm$ 2 km/s (see Table \ref{tab:list_of_analyzed_clusters}), which is a reasonable limit considering the observed dispersion by previous studies such as \cite{2009A&A...498..949M}. We did not detect any double lined spectroscopic binary stars in our dataset.

After co-addition and the second selection criteria (i.e., S/N higher than 40 and membership validation by radial velocity), the dataset is reduced to 447 spectra that correspond to 392 different stars.

\section{Spectral analysis}
\label{s:spectral_analysis}

An automatic computational process was developed to derive atmospheric parameters and chemical abundances. The process is based on the integrated spectroscopic framework named iSpec \citep{2014A&A...569A.111B}. 

For the atmospheric parameters derivation we used the atomic data kindly provided by the GES line-list sub-working group prior to publication (Heiter et al., in prep.). The line-list covers our wavelength range of interest and it also provides a selection of middle-\footnote{Lines that might be slightly more blended for hotter or colder stars} and high-quality lines (based on the reliability of the oscillator strength and the blend level) for iron and other elements (e.g., Na, Mg, Al, Si, Ca, Sc, Ti, V, Cr, Mn, Co, Ni, Cu, Zn, Sr, Y, Zr, Ba, Nd, and Sm).

We adopted the MARCS\footnote{\url{http://marcs.astro.uu.se/}} model atmosphere \citep{2008A&A...486..951G} with the solar abundances from \cite{2007SSRv..130..105G}. It is worth noting that the model atmosphere grid is formed of a combination of plane-parallel and spherical models. The first is reasonable for modeling the atmosphere of dwarf stars (where the extent of the atmosphere is smaller than the stellar radius), while the second is more appropriate for giant stars. However, the synthesizer used by iSpec \citep[SPECTRUM;][]{1994AJ....107..742G} will interpret the spherical models as plane-parallel. The differences that may be introduced for F, G, and K giants are below 0.03 dex in terms of iron abundances as shown by \cite{2006A&A...452.1039H}.

\subsection{Atmospheric parameters}
\label{sub:atmospheric_parameters_determination}

The determination of atmospheric parameters forms part of an iterative process, based on iSpec, where the continuum normalization also takes place. It consists of the following steps.

\begin{enumerate}
    \item Blind normalization. At this stage we do not know the kind of star we are analyzing, thus we fit the continuum by using the default iSpec algorithm where the following general subprocesses are executed:
        \begin{enumerate}
            \item reduction of the noise effects by applying a median filter with a window of $0.10 \mathrm{nm}$;
            \item application of a maximum filter with a window of $1.0 \mathrm{nm}$ to select those fluxes that have a larger probability to belong to the continuum;
            \item fitting of second degree splines every $1.0 \mathrm{nm}$ to the filtered points and dividing the original observed spectrum by the fitted model.
        \end{enumerate}

    \item Line fitting. For each spectrum, we fit the selected absorption lines with Gaussian profiles and we automatically discard lines that fall into one of these cases:
        \begin{enumerate}
            \item fitted Gaussian peak too far away from the expected position (more than 0.0005 nm). Convection could produce shifts, but it is also possible that a strong nearby absorption line is dominating the region and blending considerably the original targeted line. The analysis would require manual inspection, thus we reject those lines.
            \item bad fits with a root mean square bigger than 1.0 (e.g., extreme values due to a cosmic ray).
            \item absorption lines potentially affected by telluric lines (previously identified by cross-correlating with a telluric line mask).
            \item invalid fluxes (i.e., negative or inexistent due to gaps in the observation).
        \end{enumerate}
    This verification process allows us to adapt the analysis to the peculiarities of each observation, ensuring that only the best quality regions are used.

    \item Fast atmospheric parameter estimation. We use the synthetic spectral fitting technique implemented in iSpec, where a least-square algorithm compares the observed spectra with synthetic spectra. The compared regions correspond to the selected absorption lines from step 2 together with the wings of H-$\alpha$, H-$\beta$ and the Mg triplet (around 515-520~nm). The process estimates the following atmospheric parameters: effective temperature, surface gravity, metallicity, microturbulence, and macroturbulence. The rotational velocity is fixed to 2.0 km/s since it generally degenerates with the macroturbulence.
          To speed up this first estimation, we limit the minimization algorithm to one iteration and we use a small pre-computed synthetic grid with key spectra (i.e., metal-rich/poor dwarf/giant). This process allows us to quickly distinguish dwarfs from giants and overall metallicities. 

    \item Guided normalization. The same steps described in the blind normalization stage are executed, but after ignoring all the fluxes that have a value below $0.98$ in their respective synthetic spectra (computed with the fast estimation of atmospheric parameters from step 3). This way, we reduce the effect of strong lines in the normalization process.

    \item Line re-fitting. Step 2 is repeated with the new normalized spectra obtained from step 3.

    \item Final atmospheric parameter determination: The same analysis described in step 3 is repeated, but now the maximum number of iterations is increased to six, which is an optimal value as shown in \cite{2014A&A...569A.111B}.
\end{enumerate}

To reduce the dataset to mainly FGK stars in the main sequence and red giant branch, we discarded spectra for which we found an effective temperature higher than 6500 K or lower than 4500 K, and a surface gravity higher than 4.60 dex or lower than 2.00 dex (same limits as in \citealt{2014A&A...561A..93H}).

After this third selection criterion, 389 spectra remain, which correspond to 339 stars. The selection covers the 35 clusters listed in Table \ref{tab:list_of_analyzed_clusters}, where we included their coordinates, radial velocity, spectroscopic metallicity, and age from the literature.

\begin{table*}[ht!]
    \begin{center}
        \caption{List of analyzed clusters with the number of co-added spectra per instrument and other known cluster properties from the literature.}
        \label{tab:list_of_analyzed_clusters}
        \tabcolsep=0.08cm
        \begin{tabular}{l|c c c c c c c c c c c c }

    &   \scriptsize{HARPS}  &   \scriptsize{NARVAL} &   \scriptsize{UVES}   &   $l$ &   $b$ &   Distance    &   \multicolumn{2}{c}{RV}         &   \multicolumn{2}{c}{[M/H]}          &   \multicolumn{2}{c}{Age}            \\
    \textbf{Cluster}    &       &       &       &   (deg)   &   (deg)   &   (pc)    &   (km/s)  &   \scriptsize{Src} &   (dex)   &   \scriptsize{Src} &   (Gyr)   &   \scriptsize{Src} \\
        \hline                                                                                              \\
        Collinder350    &   -   &   2   &   -   &   26.75   &   14.66   &   280 &   -15.20  &   (7) &   -   &   -   &   0.4 &   (1) \\
        IC2714  &   8   &   -   &   4   &   292.40  &   -1.80   &   1240    &   -13.60  &   (11)    &   0.02    &   (9) &   0.3 &   (3) \\
        IC4651  &   7   &   -   &   29  &   340.09  &   -7.91   &   890 &   -30.36  &   (11)    &   0.12    &   (9) &   1.1 &   (1) \\
        IC4756  &   -   &   -   &   4   &   36.38   &   5.24    &   480 &   -25.16  &   (4) &   0.02    &   (9) &   0.7 &   (3) \\
        M67 &   49  &   -   &   46  &   215.70  &   31.90   &   790 &   33.77   &   (11)    &   0.00    &   (9) &   4.1 &   (3) \\
        Melotte111  &   -   &   11  &   -   &   221.35  &   84.02   &   100 &   0.01    &   (4) &   0.00    &   (9) &   0.7 &   (8) \\
        Melotte20   &   -   &   8   &   -   &   146.57  &   -5.86   &   190 &   -2.04   &   (4) &   0.14    &   (9) &   0.1 &   (8) \\
        Melotte22   &   -   &   7   &   -   &   166.57  &   -23.52  &   130 &   5.41    &   (2) &   -0.01   &   (9) &   0.1 &   (8) \\
        Melotte71   &   -   &   3   &   -   &   228.95  &   4.50    &   1970    &   50.71   &   (4) &   -0.27   &   (9) &   0.2 &   (1) \\
        NGC1817 &   -   &   4   &   -   &   186.16  &   -13.10  &   1380    &   65.31   &   (4) &   -0.11   &   (9) &   0.4 &   (1) \\
        NGC2099 &   -   &   11  &   -   &   177.64  &   3.09    &   1330    &   8.30    &   (4) &   0.02    &   (9) &   0.4 &   (8) \\
        NGC2251 &   -   &   2   &   -   &   203.58  &   0.11    &   1890    &   25.33   &   (4) &   -0.09   &   (9) &   0.3 &   (1) \\
        NGC2360 &   8   &   -   &   4   &   229.81  &   -1.43   &   770 &   27.69   &   (11)    &   -0.03   &   (9) &   0.6 &   (1) \\
        NGC2423 &   1   &   -   &   5   &   230.48  &   3.54    &   1040    &   18.47   &   (4) &   0.08    &   (9) &   1.0 &   (8) \\
        NGC2447 &   -   &   -   &   10  &   240.04  &   0.13    &   1570    &   22.08   &   (4) &   -0.03   &   (9) &   0.6 &   (8) \\
        NGC2477 &   43  &   -   &   -   &   253.56  &   -5.84   &   1360    &   7.61    &   (11)    &   0.07    &   (9) &   0.9 &   (3) \\
        NGC2539 &   7   &   -   &   5   &   233.71  &   11.11   &   360 &   29.23   &   (11)    &   -0.02   &   (9) &   0.5 &   (3) \\
        NGC2547 &   -   &   -   &   1   &   264.46  &   -8.60   &   770 &   14.00   &   (7) &   -0.16   &   (1) &   0.0 &   (5) \\
        NGC2548 &   -   &   2   &   -   &   227.87  &   15.39   &   1680    &   7.70    &   (4) &   0.08    &   (1) &   0.4 &   (3) \\
        NGC2567 &   4   &   -   &   2   &   249.80  &   2.96    &   190 &   36.71   &   (11)    &   -0.04   &   (9) &   0.3 &   (8) \\
        NGC2632 &   2   &   17  &   1   &   205.92  &   32.48   &   910 &   34.07   &   (4) &   0.20    &   (9) &   0.6 &   (8) \\
        NGC3114 &   -   &   -   &   2   &   283.33  &   -3.84   &   490 &   -1.72   &   (4) &   0.05    &   (9) &   0.2 &   (8) \\
        NGC3532 &   4   &   -   &   3   &   289.57  &   1.35    &   940 &   4.23    &   (11)    &   0.00    &   (9) &   0.4 &   (8) \\
        NGC3680 &   -   &   -   &   8   &   286.76  &   16.92   &   2180    &   1.28    &   (4) &   -0.01   &   (9) &   1.2 &   (1) \\
        NGC4349 &   5   &   -   &   4   &   299.72  &   0.83    &   930 &   -11.33  &   (11)    &   -0.07   &   (9) &   0.2 &   (1) \\
        NGC5822 &   2   &   -   &   3   &   321.57  &   3.59    &   300 &   -29.31  &   (4) &   0.08    &   (9) &   0.9 &   (1) \\
        NGC6253 &   3   &   -   &   -   &   335.46  &   -6.25   &   630 &   -29.11  &   (6) &   0.34    &   (9) &   5.0 &   (1) \\
        NGC6475 &   -   &   -   &   1   &   355.86  &   -4.50   &   380 &   -15.57  &   (4) &   0.02    &   (9) &   0.3 &   (8) \\
        NGC6494 &   4   &   -   &   1   &   9.89    &   2.83    &   1880    &   -8.08   &   (11)    &   -0.04   &   (9) &   0.3 &   (1) \\
        NGC6633 &   1   &   -   &   3   &   36.01   &   8.33    &   1220    &   -28.95  &   (4) &   -0.08   &   (9) &   0.4 &   (1) \\
        NGC6705 &   19  &   -   &   2   &   27.31   &   -2.78   &   770 &   34.70   &   (11)    &   0.12    &   (9) &   0.3 &   (8) \\
        NGC6811 &   -   &   2   &   -   &   79.21   &   12.01   &   460 &   7.28    &   (4) &   0.04    &   (10)    &   0.6 &   (3) \\
        NGC6940 &   -   &   5   &   -   &   69.86   &   -7.15   &   3300    &   7.89    &   (4) &   0.01    &   (1) &   0.7 &   (1) \\
        NGC7092 &   -   &   3   &   -   &   92.40   &   -2.24   &   11400   &   -2.80   &   (2) &   0.01    &   (1) &   0.4 &   (8) \\
        NGC752  &   -   &   7   &   -   &   137.13  &   -23.25  &   3000    &   5.04    &   (4) &   -0.02   &   (9) &   1.1 &   (5) \\

        \end{tabular}
    \end{center}
    \tablefoot{Galactic coordinates and distances from \cite{2002A&A...389..871D}. Radial velocities, spectroscopic metallicities and indicative ages from
        (1)  \citealt{2002A&A...389..871D},
        (2)  \citealt{2005A&A...438.1163K}, 
        (3)  \citealt{2006MNRAS.371.1641P}, 
        (4)  \citealt{2007A&A...476..217C}, 
        (5)  \citealt{2010A&A...514A..81P}, 
        (6)  \citealt{2011A&A...535A..39M}, 
        (7)  \citealt{2012AstL...38..331A}, 
        (8)  \citealt{2013A&A...557A..10N},
        (9)  \citealt{2014A&A...561A..93H},
        (10) \citealt{2014MNRAS.445.2446M},
        and (11) HARPS.
    }
\end{table*}

\subsection{Chemical abundances}
\label{s:chemical_abundances}

The metallicity obtained in the atmospheric parameter determination process (Sect.~\ref{sub:atmospheric_parameters_determination}) corresponds to a global scaling factor that is applied to all the elements (taking the solar abundance as the reference point). In a subsequent step, individual abundances are derived.

It has been shown that line-by-line differential analysis with a manual selection of lines can reach precision of 0.003 dex for solar twins \citep{2012A&A...543A..29M}. In our case, the whole analysis is automatized and stars cover a wider parameter space, thus we cannot expect such a level of precision. Nevertheless, we developed a differential analysis where we have derived the solar absolute abundances of all the absorption lines that were selected for 7 solar spectra included in the Gaia FGK Benchmark Stars library  \citep{2014A&A...566A..98B}. The process follows these steps for each element and spectrum:

\begin{enumerate}
    \item derivation of the absolute abundances for each of the selected lines by using the synthetic spectral technique implemented in iSpec.
    \item calculation of the relative abundances by subtracting the solar absolute abundances from the absolute abundances for each line.
    \item derivation of the final abundance for the spectrum by calculating the weighted average
        \begin{equation}
            \bar{x}_\mathrm{w} = \frac{ \sum_{i=1}^n w_i x_i}{\sum_{i=1}^n w_i},
        \end{equation}
        where the weight is the inverse of the abundance error $w = \frac{1}{eA(X)^2}$ reported by iSpec, which is influenced by the spectral S/N and the goodness of fit.
    \item derivation of the error associated with the final abundance from the unbiased weighted sample dispersion
        \begin{equation}
            \sigma_\mathrm{w} = \frac{V_1}{V_1^2 - V_2}  \sqrt{\frac{\sum_{i=1}^n w_i \left(x_i - \bar{x}_\mathrm{w} \right)^2 }{\sum_{i=1}^n w_i}},
        \end{equation}
        where $V_1 = \sum_{i=1}^{n} w_i$ and $V_2 = \sum_{i=1}^{n} w^2_i$. 
\end{enumerate}

We analyzed 779,422 lines from which we discarded extreme values which are clearly outliers (i.e., relative abundances bigger than 1.0 dex and smaller than -5.0 dex) and elements with only one measured line (we required at least two lines to be able to calculate the weighted sample dispersion). As an order of magnitude, the typical dispersion of the abundance of a single element is $\sim$0.10 dex.

The abundances that were successfully measured in all the spectra cover 17 species corresponding to 14 different elements. The list includes iron peak elements (V I, Cr I/II, Mn I, Fe I/II, Co I, Ni I), alpha elements (Mg I, Si I, Ca I, Ti I/II), odd-Z elements (Na I, Sc II), and s-process elements (Ba II, Y II). We discarded stars for which we do not have any of the previous abundances, thus the final dataset was reduced to 206 stars covering 32 clusters.

\subsection{Chemical outliers}
\label{sub:chemical_outliers}

Assuming that stars born together share the same chemical signature, we used the 17 abundances to identify outliers in each cluster. However, the low number of stars per cluster is a limiting factor for the detection of outliers. To overcome this limitation, we first compressed the 17 dimensions (abundance values) by using the statistical procedure named principal component analysis (PCA), where the abundances are converted into a set of linearly uncorrelated variables (named components) by applying an orthogonal transformation. The first principal component has the largest possible variance, and each subsequent component has the highest variance possible under the constraint that it is orthogonal to (i.e., uncorrelated with) the preceding components. \cite{2012MNRAS.421.1231T} illustrated the power of PCA to study and interpret stellar element abundances.

Second, using the first two components, we estimated and subtracted the central location in the PCA space for each independent cluster. Thus, all the clusters share the same central location (i.e., origin of coordinates in the PCA space) and outlying stars are placed farther away. It is important to minimize the impact of deviant values in the determination of clusters' central locations in the PCA space. Therefore, we used a robust estimator named minimum covariance determinant (MCD) and the Mahalanobis distance \citep{mahalanobis1936generalized}, which can tolerate the effect of nearly 50\% of contamination in the data \citep{rousseeuw1984least} and for which there is a computationally fast and well-known algorithm \citep{rousseeuw1999fast}. 

\begin{figure*}
    \begin{centering}
        \includegraphics[width=9cm, trim = 1mm 1mm 1mm 1mm, clip]{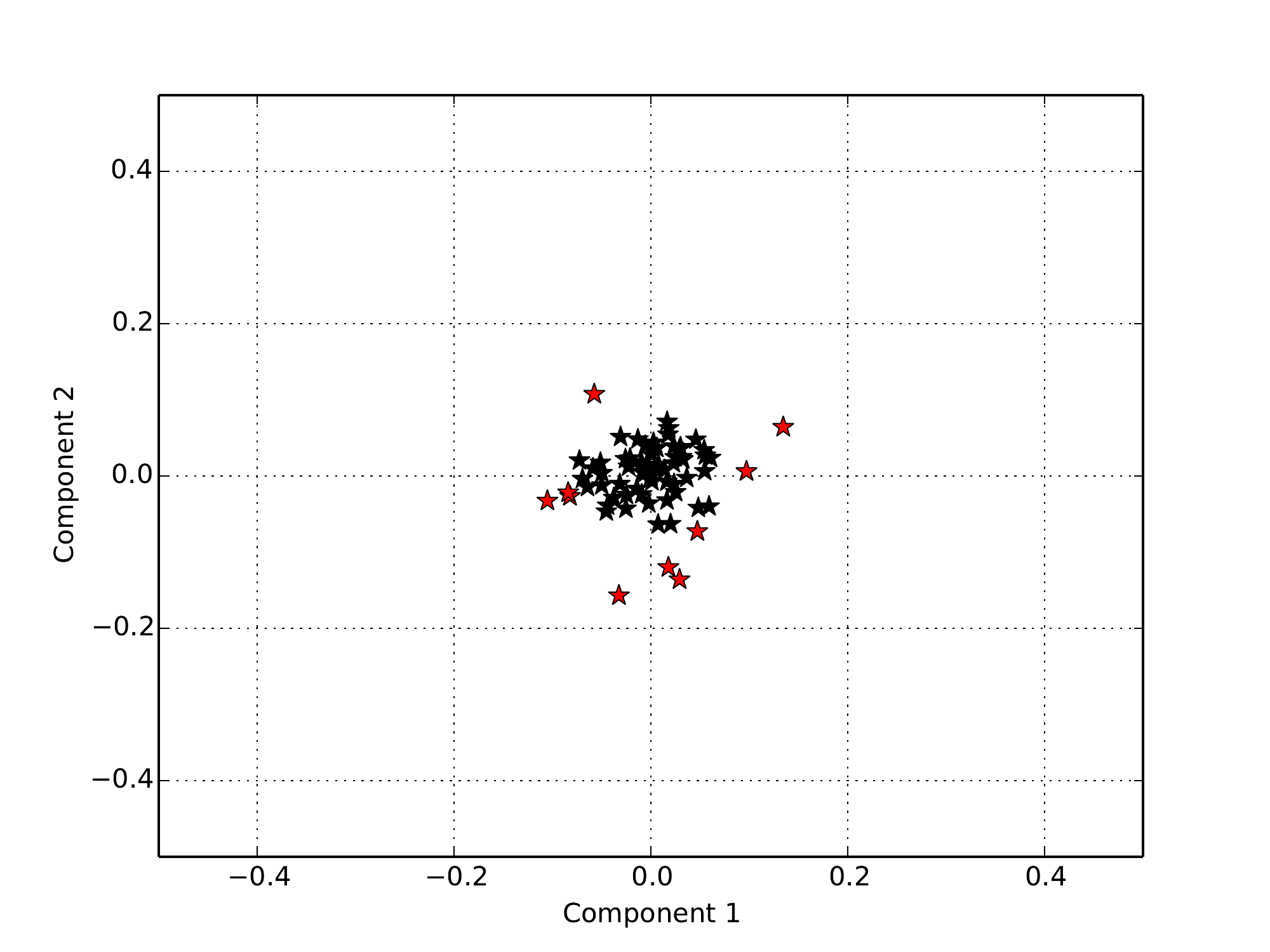}
        \includegraphics[width=9cm, trim = 1mm 1mm 1mm 1mm, clip]{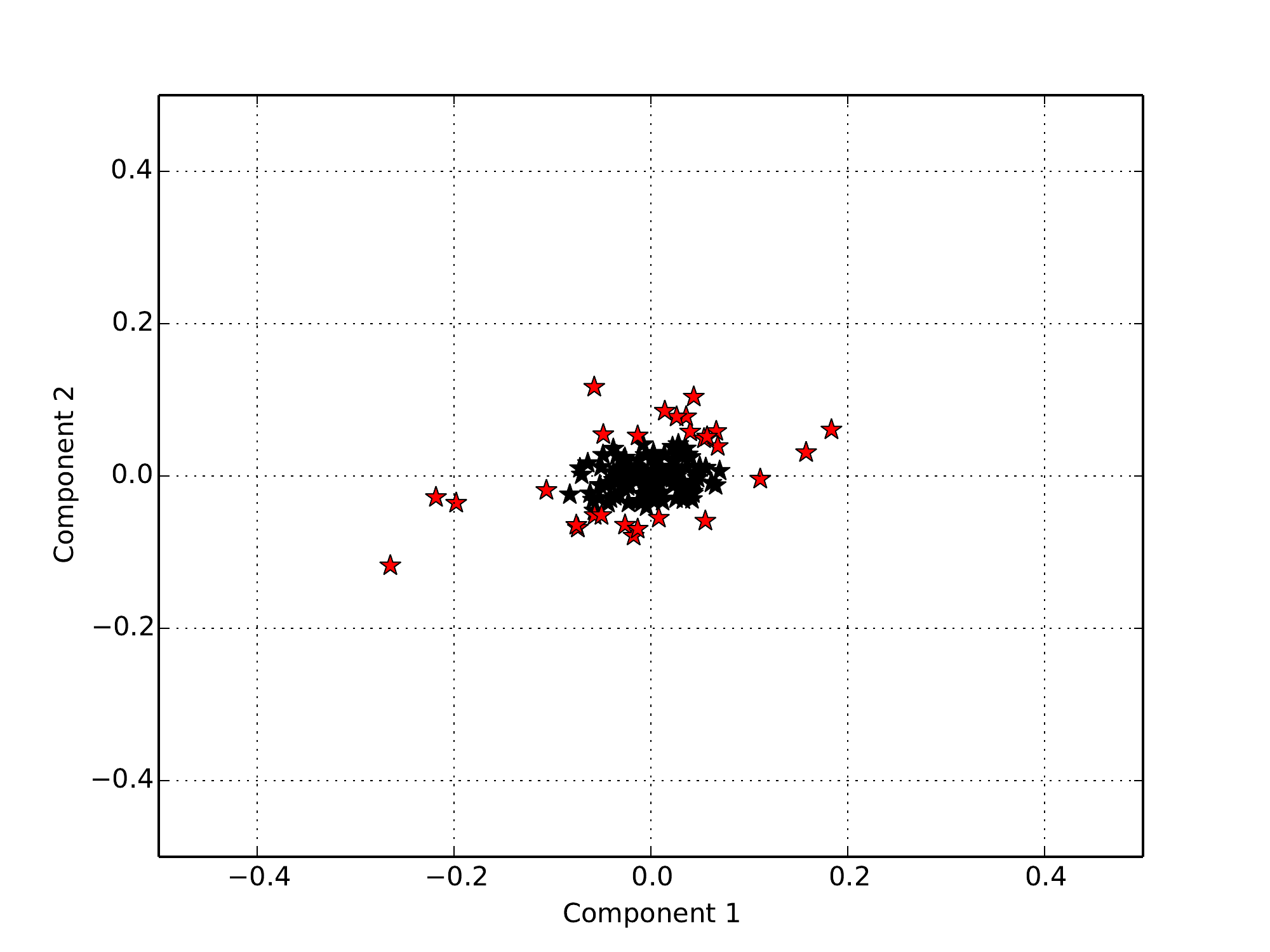}
        \par
    \end{centering}
    \caption{Dwarf (left) and giant stars (right) from all clusters represented using the first two components of the PCA with the central cluster location subtracted. Stars chemically identified as outliers are shown in red.}
    \label{fig:outliers_pca}
\end{figure*}

Finally, we redetermined the central location and scatter of all the stars applying again the same algorithm and assuming that all the cluster have similar dispersion. For multivariate normally distributed data, the Mahalanobis distances to the central location are approximately chi-square distributed with $p$ degrees of freedom \citep[$\chi^{2}_{p}$ where $p=2$ in our case since we use the first two principal components;][]{filzmoser2004multivariate}. We tagged as outliers those stars with a squared distance higher than the 80\% quantile of the chi-squared distribution (see Fig. \ref{fig:outliers_pca}).

\begin{figure*}
    \begin{centering}
        \includegraphics[width=9cm, trim = 1mm 1mm 1mm 1mm, clip]{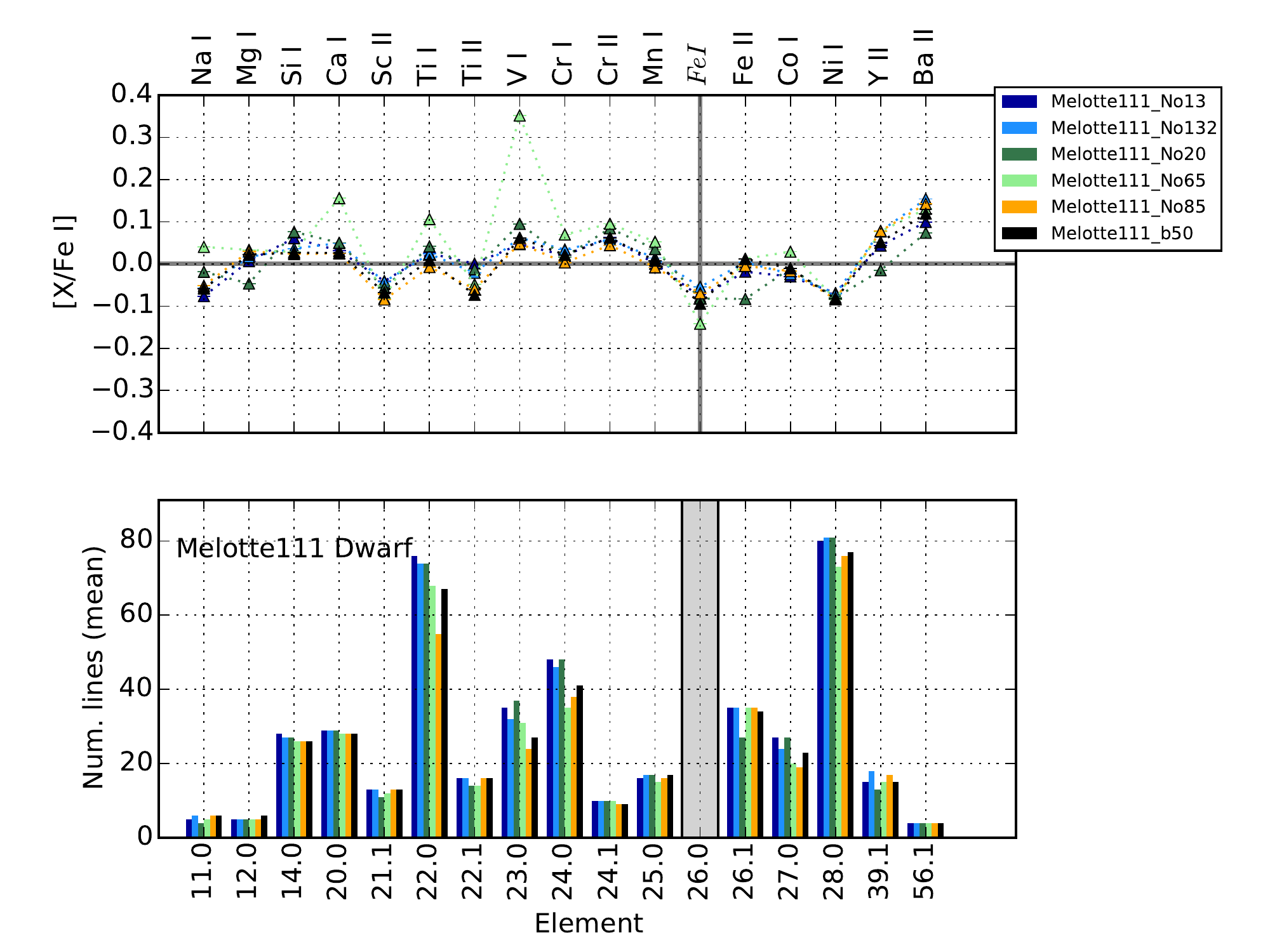}
        \includegraphics[width=9cm, trim = 1mm 1mm 1mm 1mm, clip]{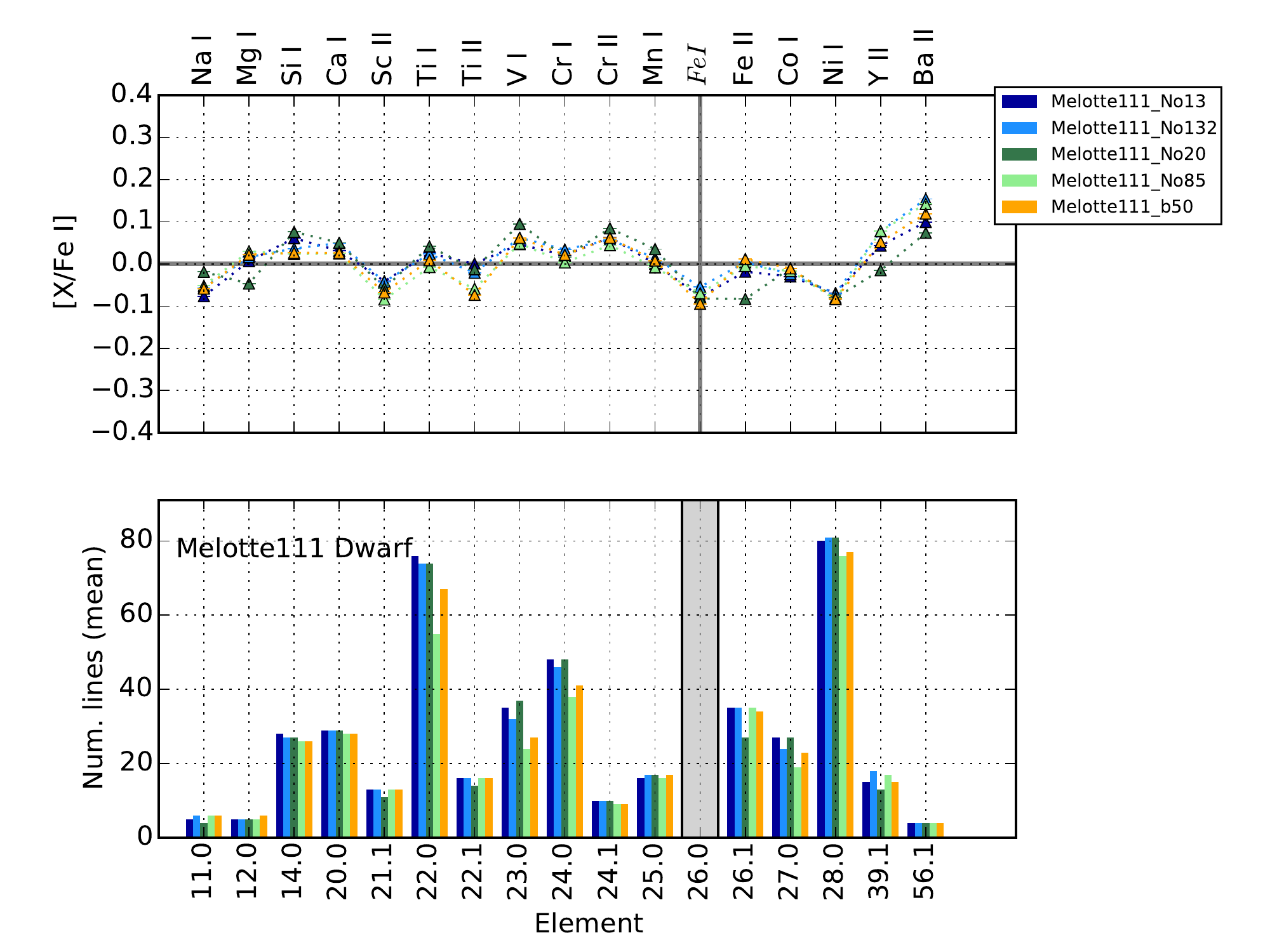}
        \par
    \end{centering}
    \caption{Abundances (top) and mean number of lines used (bottom) in function of species (element code at top, atomic number, and ionization state in Kurucz format at bottom where '0' is neutral and '1' is ionized) for Melotte 111 dwarfs showing all the analyzed stars (left) and after filtering outliers (right). All the abundance ratios are referenced to iron except iron itself, which is relative to hydrogen. Each color represents a star with an identification name shown in the legend.}
    \label{fig:abundance_outliers}
\end{figure*}

After executing this procedure and discarding the identified outliers (see Fig.~\ref{fig:abundance_outliers}), the dataset was reduced to 177 stars corresponding to 31 clusters. The discarded stars could be chemically peculiar or simply not belong to the cluster. In the case of NGC3114, the two stars of the cluster were classified as outliers by this method, probably only one of them is a real outlier, but because of the lack of statistics (if there is an outlier in a cluster of only two stars, the mean abundances are strongly affected) we prefer to be conservative and discard the complete cluster. We decided not to perform any further detailed analysis of these stars since this would fall outside of the scope of this work. Detailed abundances per cluster and stellar type can be found in Tables \ref{tab:abundances_startype} and \ref{tab:abundances_cluster}.

\section{Chemical tagging}
\label{s:chemical_tagging}

\subsection{Continuum normalization effects}

As pointed out in the introduction, some of the chemical studies found in the literature are based on non-homogeneous compilations of abundances, obtained from different sources and by different methods (e.g., equivalent width/synthetic spectra) and ingredients (e.g., atomic data, model atmospheres). This inhomogeneity implies systematic uncertainties that can mislead our scientific conclusions.

To illustrate the impact of those changes on the metallicity, we repeated our full analysis changing only one parameter in the continuum normalization process (see Section \ref{sub:atmospheric_parameters_determination}). We decreased the median filter window from $0.10$ to $0.01~\mathrm{nm}$, which raises the continuum placement.

We used the iron abundance [Fe/H] as a pertinent tracer of the metallicity, and we compared the values obtained from both slightly different normalization procedures. A small change in the normalization criteria produces a systematic average difference of $-0.07\pm0.04$ dex.

When we compared both results (median filter window of $0.10$ and $0.01 \mathrm{nm}$) with open cluster metallicities found in the literature, we obtained the average differences of $-0.07\pm0.07$ dex and $0.01\pm0.08$ dex, respectively (details in Table \ref{tab:metallicities_vs_continuum}). The large dispersion confirms the inadequacy of mixing chemical abundances from different literature sources to draw solid scientific conclusions.

\begin{table}[ht!]
    \begin{center}
        \caption{Iron abundances from neutral lines when using two slightly different normalization processes, and comparison to literature metallicities.}
        \label{tab:metallicities_vs_continuum}
        \tabcolsep=0.10cm
        \begin{tabular}{l|c c | c | c c | c | c | c }

                  & \multicolumn{3}{c|}{[Fe I/H]$_{1}$} & \multicolumn{3}{c|}{[Fe I/H]$_{2}$} & \multirow{2}{*}{$\Delta$\scriptsize{Cont}} & \multirow{2}{*}{$\star$} \\
\textbf{cluster}  &   $\bar{x}_\mathrm{w}$   &   $\sigma_\mathrm{w}$ & $\Delta$\scriptsize{Lit}  &   $\bar{x}_\mathrm{w}$   &   $\sigma_\mathrm{w}$ & $\Delta$\scriptsize{Lit}   &    &   \\
        \hline

Collinder350    &   -0.10   &   -   &   -   &   0.00    &   -   &   -   &   -0.10   &   1   \\
IC2714  &   -0.12   &   0.02    &   -0.14   &   -0.08   &   0.10    &   -0.10   &   -0.04   &   6   \\
IC4651  &   0.00    &   0.03    &   -0.12   &   0.06    &   0.16    &   -0.06   &   -0.06   &   8   \\
IC4756  &   -0.09   &   0.04    &   -0.11   &   0.00    &   0.07    &   -0.02   &   -0.09   &   4   \\
M67 &   -0.06   &   0.04    &   -0.06   &   0.00    &   0.13    &   0.00    &   -0.06   &   42  \\
Melotte111  &   -0.08   &   0.01    &   -0.08   &   0.00    &   0.09    &   0.00    &   -0.08   &   5   \\
Melotte20   &   0.03    &   -   &   -0.11   &   0.18    &   -   &   0.04    &   -0.15   &   1   \\
Melotte22   &   -0.06   &   -   &   -0.05   &   0.09    &   -   &   0.10    &   -0.15   &   1   \\
Melotte71   &   -0.09   &   -   &   0.18    &   -0.02   &   -   &   0.25    &   -0.07   &   1   \\
NGC1817 &   -0.23   &   0.02    &   -0.12   &   -0.11   &   0.02    &   0.00    &   -0.12   &   3   \\
NGC2099 &   -0.03   &   -   &   -0.05   &   0.11    &   -   &   0.09    &   -0.14   &   1   \\
NGC2251 &   -0.09   &   -   &   0.00    &   -0.01   &   -   &   0.08    &   -0.08   &   1   \\
NGC2360 &   -0.10   &   0.03    &   -0.07   &   -0.07   &   0.09    &   -0.04   &   -0.03   &   7   \\
NGC2423 &   0.02    &   0.01    &   -0.06   &   0.09    &   0.07    &   0.01    &   -0.07   &   2   \\
NGC2447 &   -0.13   &   0.02    &   -0.10   &   -0.03   &   0.09    &   0.00    &   -0.10   &   6   \\
NGC2477 &   -0.02   &   0.02    &   -0.09   &   0.02    &   0.08    &   -0.05   &   -0.04   &   29  \\
NGC2539 &   -0.04   &   0.03    &   -0.02   &   0.00    &   0.10    &   0.02    &   -0.04   &   7   \\
NGC2547 &   -0.13   &   -   &   0.03    &   -0.06   &   -   &   0.10    &   -0.07   &   1   \\
NGC2567 &   -0.14   &   0.01    &   -0.10   &   -0.09   &   0.09    &   -0.05   &   -0.05   &   4   \\
NGC2632 &   0.10    &   0.03    &   -0.10   &   0.20    &   0.11    &   0.00    &   -0.10   &   13  \\
NGC3532 &   -0.12   &   0.03    &   -0.12   &   -0.06   &   0.09    &   -0.06   &   -0.06   &   3   \\
NGC3680 &   -0.12   &   0.04    &   -0.11   &   -0.06   &   0.13    &   -0.05   &   -0.06   &   5   \\

NGC4349 &   -0.16   &   0.03    &   -0.09   &   -0.12   &   0.10    &   -0.05   &   -0.04   &   3   \\
NGC5822 &   -0.12   &   0.01    &   -0.20   &   -0.02   &   0.09    &   -0.10   &   -0.10   &   2   \\
NGC6475 &   0.04    &   -   &   0.02    &   0.15    &   -   &   0.13    &   -0.11   &   1   \\
NGC6494 &   -0.12   &   0.01    &   -0.08   &   -0.06   &   0.09    &   -0.02   &   -0.06   &   4   \\
NGC6633 &   -0.08   &   0.02    &   0.00    &   0.01    &   0.08    &   0.09    &   -0.09   &   4   \\
NGC6705 &   -0.01   &   0.01    &   -0.13   &   -0.04   &   0.11    &   -0.16   &   0.03    &   2   \\
NGC6811 &   -0.10   &   0.00    &   -0.14   &   -0.02   &   0.07    &   -0.06   &   -0.08   &   2   \\
NGC6940 &   0.04    &   -   &   0.03    &   0.09    &   0.07    &   0.08    &   -0.05   &   1   \\
NGC752  &   -0.07   &   0.01    &   -0.05   &   -0.02   &   0.07    &   0.00    &   -0.05   &   7   \\

        \end{tabular}
    \end{center}
    \tablefoot{ [Fe I/H]$_{1}$ normalized with a median filter window of $0.10$ nm, [Fe I/H]$_{2}$ normalized with a window of $0.01$ nm, $\Delta$Lit corresponds to the difference with literature metallicities (see Table \ref{tab:list_of_analyzed_clusters}), $\Delta$Cont represents the difference between [Fe I/H]$_{1}$ and [Fe I/H]$_{2}$. The number of stars is indicated in the last column ($\star$). }
\end{table}

For the rest of the abundances, we chose to work in terms of [X/Fe]\footnote{By definition, $\mbox{[X/Y]} \equiv \log_{10} (\mbox{N}_{X}/\mbox{N}_{Y})_{\mbox{star}} - \log_{10} (\mbox{N}_{X}/\mbox{N}_{Y})_{\odot}$, where $\mbox{N}_{X}$ and $\mbox{N}_{Y}$ are the abundances of element $X$ and element $Y$ respectively.} since this way we can partially cancel out the effect of different continuum placements.

\subsection{Abundance and astrophysical parameter correlations}

Previous works have already detected trends in chemical abundances with effective temperature and surface gravity for metal poor globular clusters, measuring iron abundances smaller in the turnoff stars than in the red giants on the order of the order of 30\% or 0.13 dex \citep{2007ApJ...671..402K, 2008A&A...490..777L, 2012ApJ...753...48N, 2013A&A...555A..31G}.

Systematic differences were also found for open clusters with metallicities closer to solar such as M67 \citep[][]{2014A&A...562A.102O}, NGC5822, or IC4756 \citep[differences higher than 0.15 dex for Na, Si, and Ti;][]{2010A&A...515A..28P}. 

The assumption of local thermodynamic equilibrium (LTE) may introduce systematic errors and abundance trends when analyzing stars covering large intervals of effective temperatures, surface gravities and metallicities. For instance, \cite{2006A&A...450..557R} exposes that the Na difference between dwarfs and giants in M67 can be explained by NLTE effects that are larger for cool giants than for warm dwarfs \citep{2000ARep...44..790M}.

The depletion of some elements could also be caused by the atomic diffusion (pushing heavier elements in the direction of increasing pressure and temperature) that takes place during the main sequence lifetime of the star and modifies the chemical composition of the stellar atmosphere. The effect is 
element-specific since radiative levitation reduces the gravitational acceleration (caused by the interaction of photons with gas particles) and acts selectively on different atoms and ions. When the star evolves toward the red giant branch, elements previously drained from the surface are mixed up again as the outer convection zone gradually reaches deeper layers. 

Additionally, as shown in \cite{2014A&A...569A.111B}, when simultaneously deriving the effective temperature, surface gravity and metallicity from spectra, degeneracies among those parameters lead to correlations where lower metallicities are found for lower surface gravities and vice versa.

\begin{figure}
    \begin{centering}
        \includegraphics[width=\linewidth, trim = 1mm 1mm 1mm 1mm, clip]{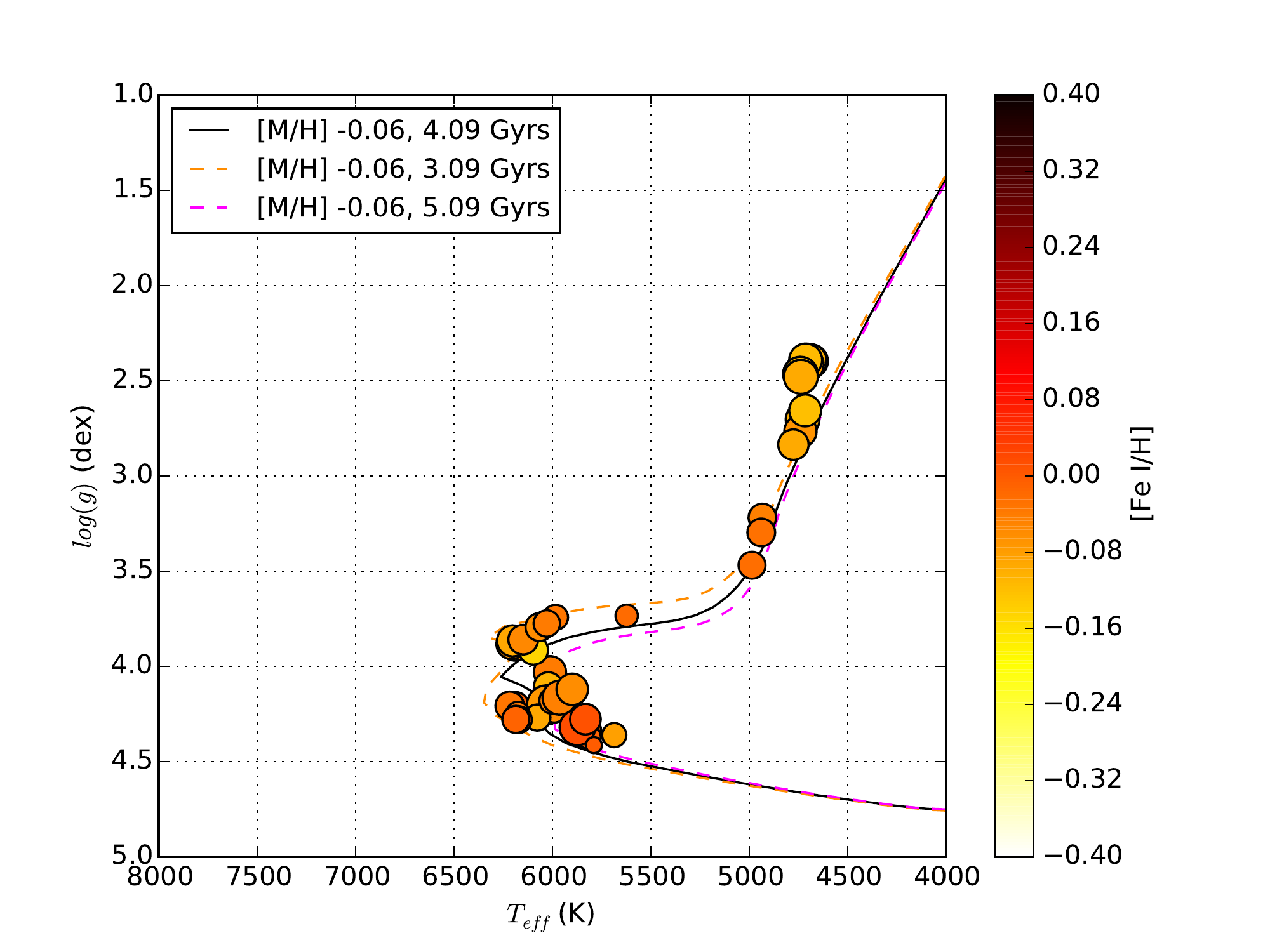}
        \par
    \end{centering}
    \caption{Hertzsprung–Russell diagram for M67 with Yonsei-Yale isochrones \citep[three different ages;][]{2004ApJS..155..667D}, color scale corresponding to the neutral iron abundance, and size represents abundance dispersion.}
    \label{fig:M67_HR_fe}
\end{figure}

\begin{figure}
    \begin{centering}
        \includegraphics[width=\linewidth, trim = 1mm 1mm 1mm 1mm, clip]{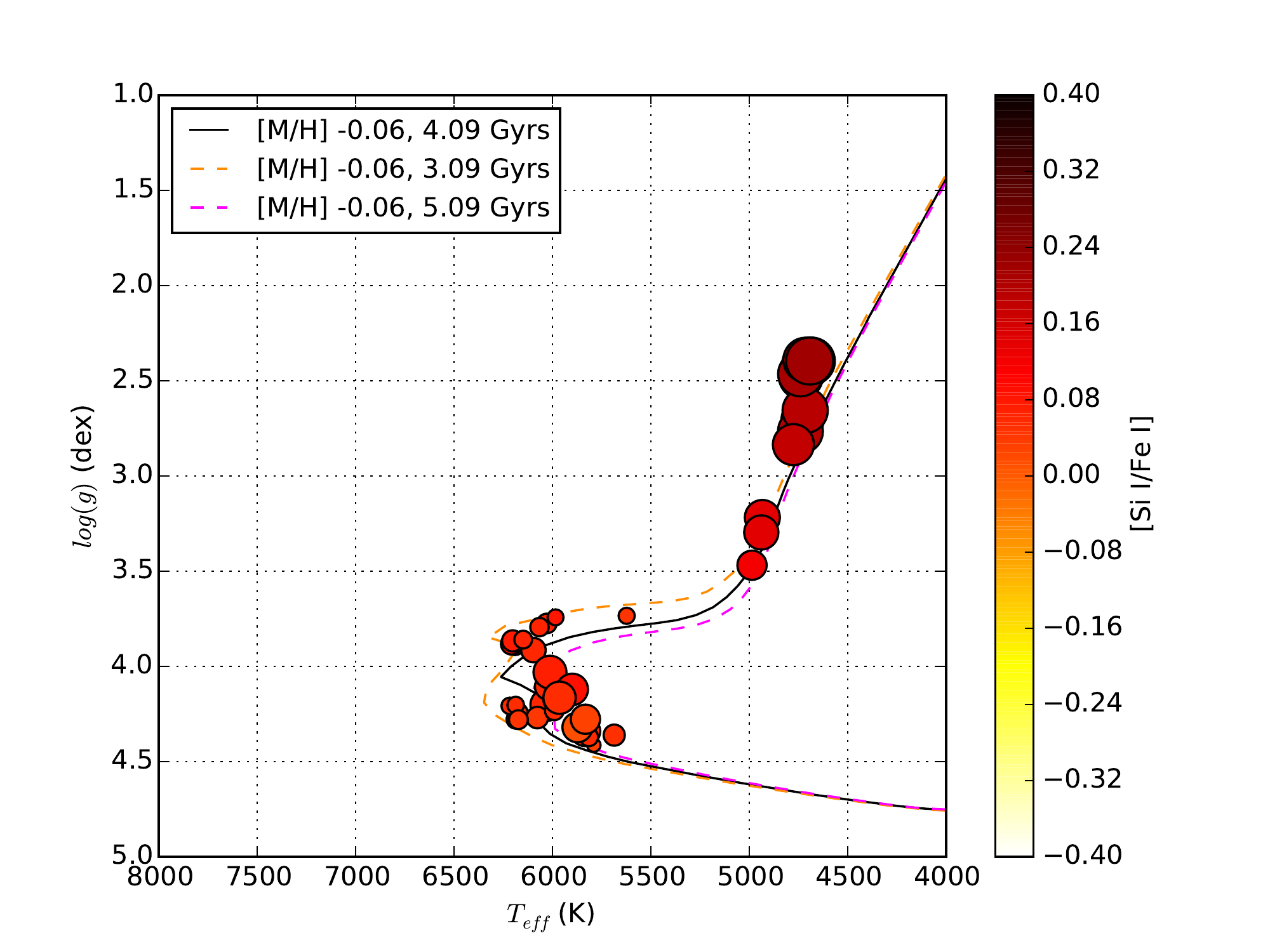}
        \par
    \end{centering}
    \caption{Hertzsprung–Russell diagram for M67 with Yonsei-Yale isochrones \citep[three different ages;][]{2004ApJS..155..667D}, color scale corresponding to the silicon abundance, and size represents abundance dispersion.}
    \label{fig:M67_HR_si}
\end{figure}

Consequently, chemical abundances derived for stars in different evolutionary stages might be affected by NLTE effects, atomic diffusion processes, and correlations from atmospheric parameter determinations. The NLTE effects can be partly canceled out only for solar dwarfs by performing differential analysis (see Sect. \ref{s:chemical_abundances}), and the effects from parameter determinations can be reduced by working with [X/Fe] ratios\footnote{[X/Fe] = [X/H] - [Fe/H]}. Additionally, to control these effects in our work, we decided to divide each cluster into two subgroups formed of dwarfs and giant stars (i.e., log(g) $\leq$ 3.5 dex).

It is worth noting that the outlier detection process (see Sect.~\ref{sub:chemical_outliers}) was executed at this subgroup level for each cluster, otherwise we would have detected a significant number of false outliers due to these stellar processes.

\begin{figure*}
    \begin{centering}
        \includegraphics[width=9cm, trim = 1mm 1mm 1mm 1mm, clip]{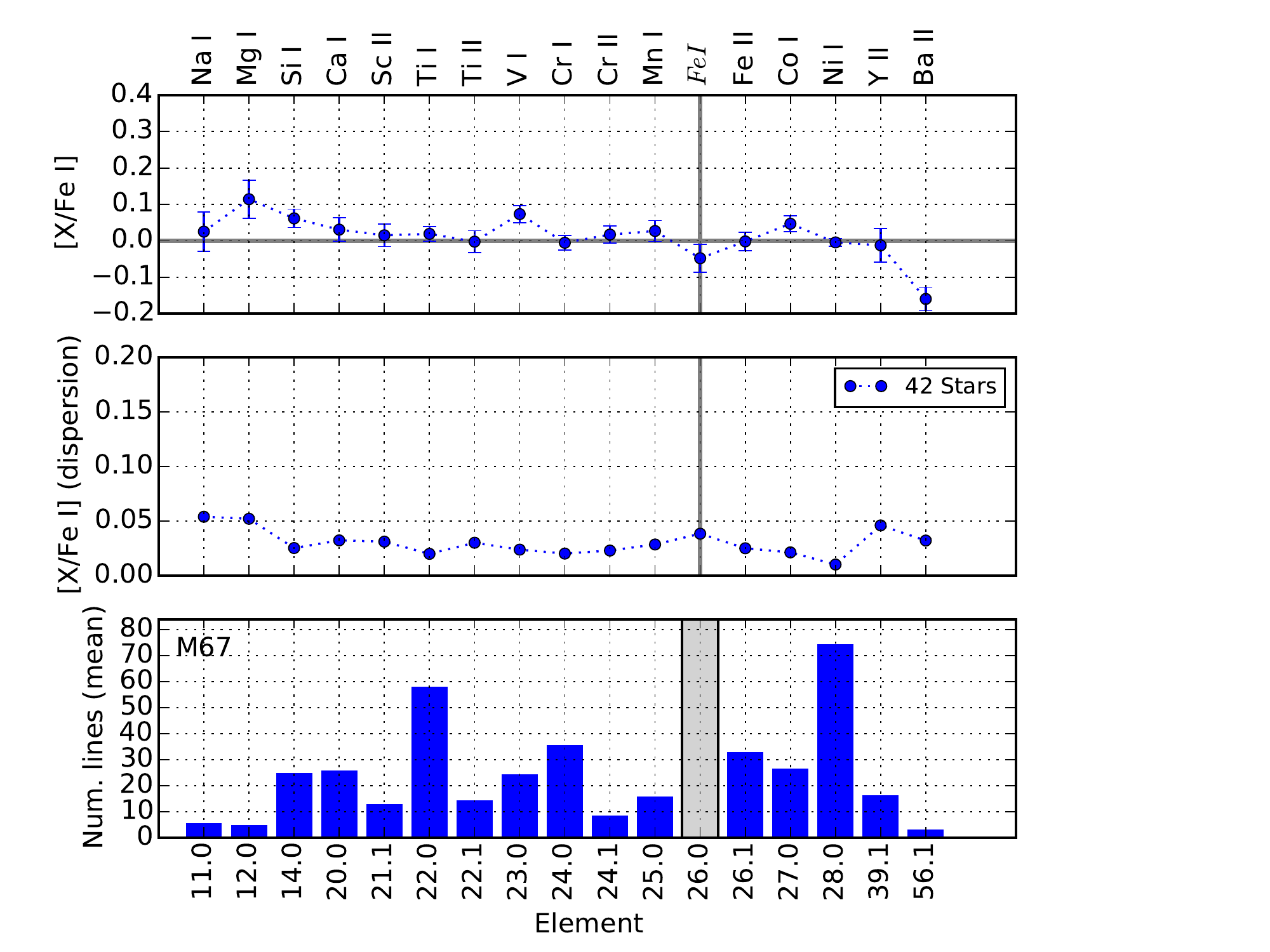}
        \includegraphics[width=9cm, trim = 1mm 1mm 1mm 1mm, clip]{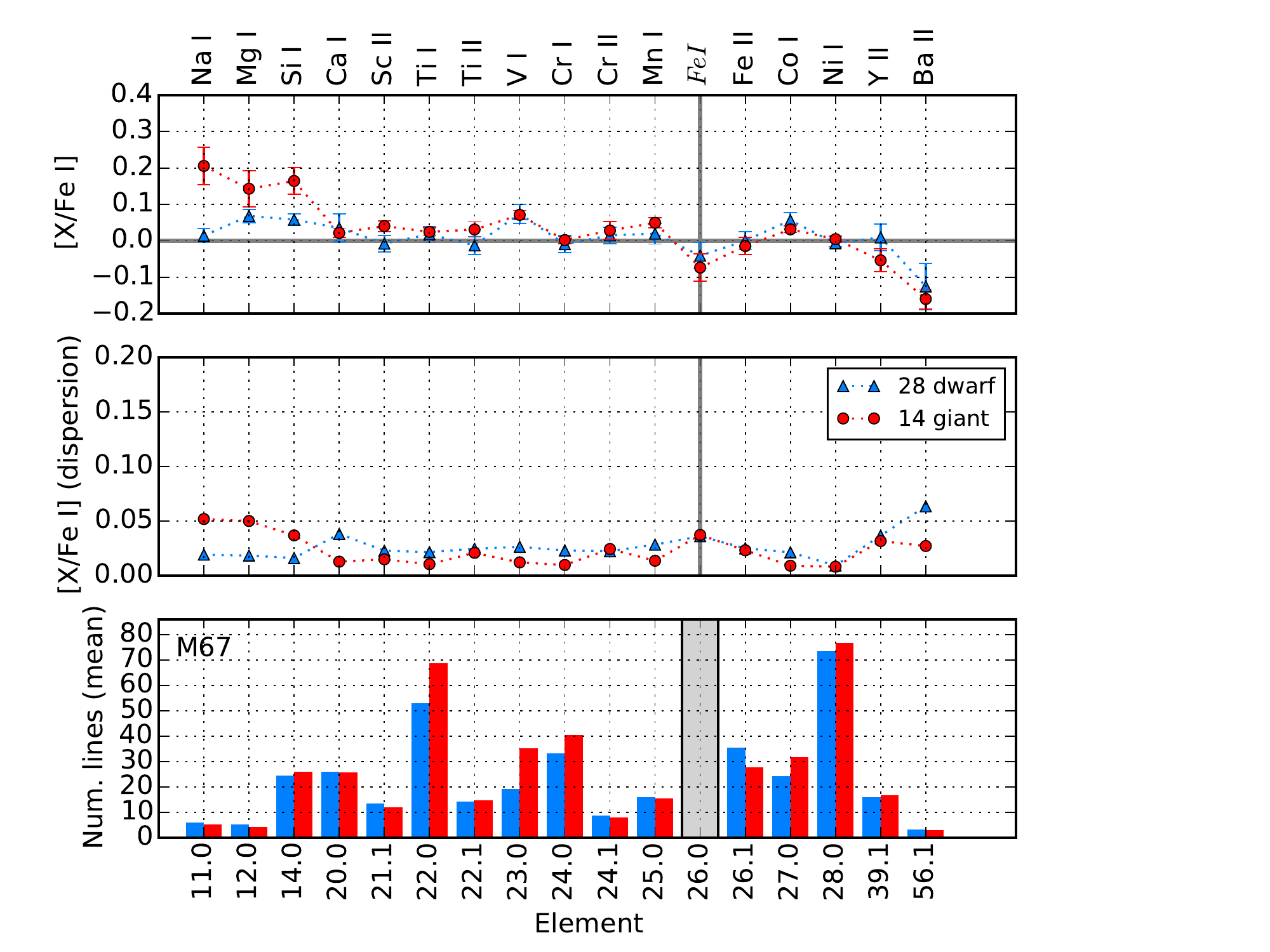}
        \par
    \end{centering}
    \caption{Average chemical abundances (top), dispersion (middle), and mean number of lines (bottom) used for M67 stars (left) and divided into two stellar-type groups (right). All the abundance ratios are referenced to iron except iron itself, which is relative to hydrogen.}
    \label{fig:M67_subgroups}
\end{figure*}

In Fig. \ref{fig:M67_subgroups} we present the chemical pattern for M67, one of the clusters with a high number of spectra in our dataset. The signature is different for each subgroup and the elemental abundance dispersion is slightly lower when we subdivide clusters per evolutionary stages. 

The chemical differences for IC4651, M67, NGC2447, NGC2632, and NGC3680 stars in various evolutionary stages for each analyzed element are shown in Table \ref{tab:chemical_diff_evolutionary_stages}. We observe that Si I, and Na I are enhanced for evolved stars with increases higher than 0.10 dex. We priorize abundances for neutral elements when possible since we have more lines for them. The rest present smaller variations that in most of the cases fall inside the error margins ($\sim$0.05 dex). For a better visual inspection, the iron and silicon abundances for M67 are shown in Fig. \ref{fig:M67_HR_fe} and Fig. \ref{fig:M67_HR_si}, respectively.

\begin{table*}[ht!]
    \begin{center}
        \caption{Chemical differences for stars in different evolutionary stages.}
        \label{tab:chemical_diff_evolutionary_stages}
        \tabcolsep=0.15cm
        \begin{tabular}{l l | c c | c c | c c | c c | c c | c c | }

    &               &   \multicolumn{2}{c|}{IC4651}         &                   \multicolumn{2}{c|}{M67}            &   \multicolumn{2}{c|}{NGC2447}            &   \multicolumn{2}{c|}{NGC2632}            &   \multicolumn{2}{c|}{NGC3680}            \\
    &               &   \scriptsize{D}  &   \scriptsize{$\Delta$G}  &                   \scriptsize{D}  &   \scriptsize{$\Delta$G}  &   \scriptsize{D}  &   \scriptsize{$\Delta$G}  &   \scriptsize{D}  &   \scriptsize{$\Delta$G}  &   \scriptsize{D}  &   \scriptsize{$\Delta$G}  \\
        \hline                                                                                                          
\parbox[t]{2mm}{\multirow{8}{*}{\rotatebox[origin=c]{90}{Iron peak}}}   &   [Fe I/H]            &   0.02    &   -0.04   &                   -0.04   &   -0.03   &   -0.12   &   -0.01   &   0.11    &   -0.04   &   -0.10   &   -0.03   \\
    &   [Fe II/Fe]          &   0.04    &   -0.07   &                   0.00    &   -0.01   &   0.00    &   -0.01   &   0.05    &   -0.05   &   0.01    &   -0.02   \\
    &   [V I/Fe]            &   0.06    &   0.01    &                   0.07    &   0.00    &   0.06    &   -0.02   &   0.07    &   -0.03   &   0.09    &   -0.05   \\
    &   [Cr I/Fe]           &   0.00    &   0.02    &                   -0.01   &   0.01    &   0.01    &   0.01    &   0.03    &   0.01    &   0.01    &   -0.01   \\
    &   [Cr II/Fe]          &   0.07    &   -0.05   &                   0.01    &   0.02    &   0.04    &   -0.03   &   0.05    &   -0.01   &   0.06    &   -0.04   \\
    &   [Mn I/Fe]           &   0.04    &   0.03    &                   0.02    &   0.03    &   -0.04   &   0.04    &   0.06    &   -0.02   &   -0.02   &   0.02    \\
    &   [Co I/Fe]           &   0.04    &   -0.01   &                   0.06    &   -0.03   &   0.00    &   -0.01   &   0.02    &   -0.01   &   0.01    &   -0.02   \\
    &   [Ni I/Fe]           &   -0.01   &   0.00    &                   -0.01   &   0.01    &   -0.09   &   0.04    &   -0.02   &   0.03    &   -0.06   &   0.02    \\
        \hline                                                                                                          
                                                                                                                    
\parbox[t]{2mm}{\multirow{5}{*}{\rotatebox[origin=c]{90}{$\alpha$ elements}}}   &   [Mg I/Fe]           &   0.06    &   0.01    &                   0.07    &   0.07    &   0.02    &   0.08    &   -0.01   &   0.05    &   0.07    &   0.07    \\
    &   [Si I/Fe]           &   0.08    &   0.09    &                   0.06    &   0.10    &   0.03    &   0.07    &   0.05    &   0.13    &   0.05    &   0.12    \\
    &   [Ca I/Fe]           &   0.05    &   -0.05   &                   0.04    &   -0.02   &   0.05    &   0.03    &   0.03    &   0.01    &   0.07    &   -0.05   \\
    &   [Ti I/Fe]           &   0.00    &   0.02    &                   0.02    &   0.00    &   0.01    &   0.04    &   -0.02   &   0.00    &   0.01    &   0.01    \\
    &   [Ti II/Fe]          &   -0.04   &   0.07    &                   -0.01   &   0.04    &   -0.04   &   0.06    &   -0.07   &   0.06    &   -0.02   &   0.06    \\
        \hline                                                                                                          
                                                                                                                    
\parbox[t]{2mm}{\multirow{4}{*}{\rotatebox[origin=c]{90}{Others}}}  &   [Na I/Fe]           &   0.01    &   0.29    &                   0.01    &   0.20    &   -0.09   &   0.33    &   0.03    &   0.31    &   -0.03   &   0.18    \\
    &   [Sc II/Fe]          &   -0.04   &   0.08    &                   -0.01   &   0.05    &   -0.07   &   0.07    &   -0.05   &   0.09    &   -0.04   &   0.08    \\

    &   [Ba II/Fe]          &   -0.04   &   0.01    &                   -0.13   &   -0.03   &   0.25    &   -0.05   &   -0.09   &   0.08    &   0.00    &   0.08    \\
    &   [Y II/Fe]           &   0.01    &   -0.01   &                   0.01    &   -0.06   &   0.10    &   0.00    &   -0.01   &   0.03    &   0.07    &   -0.04   \\
\multicolumn{2}{r|}{$\star$}                    &   6   &   2   &                   28  &   14  &   3   &   3   &   11  &   2   &   2   &   3   \\

        \end{tabular}
    \end{center}
    \tablefoot{Dwarfs (D) values and differences relative to giants (G) where G = D + $\Delta$G. Iron abundances from neutral lines ([Fe I/H]) were used as a proxy for total iron abundance ([Fe/H]). The number of stars is indicated in the last row ($\star$).}
\end{table*}

\begin{figure}
    \begin{centering}
        \includegraphics[width=\linewidth, trim = 1mm 1mm 1mm 1mm, clip]{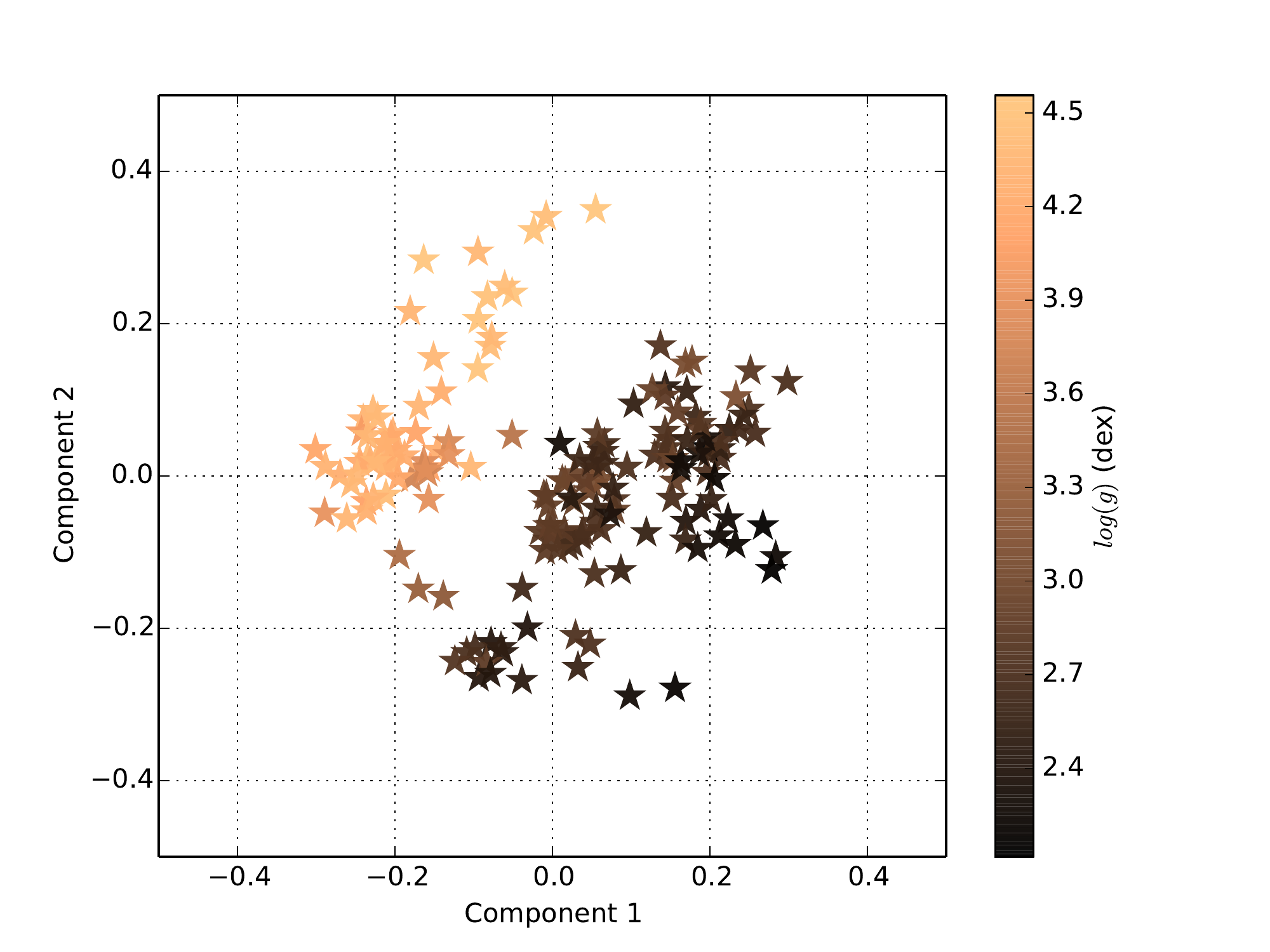}
        \par
    \end{centering}
    \caption{Abundances represented by the first two principal components. The surface gravity of each star is indicated by the color scale bar.}
    \label{fig:Clusters_PCA}
\end{figure}

An alternative approach to visually evaluating the chemical differences between subgroups is to reduce the 17 dimensions to two components using PCA as shown in Fig. \ref{fig:Clusters_PCA}. Dwarfs reside in a clearly different parameter space from giants, showing that the subgroups have distinct chemical patterns.

Regardless of whether these abundance enhancements are real or due to systematic errors (e.g., model assumptions, data treatment), they have implications that we cannot ignore for the chemical tagging technique when applied to stars in very different evolutionary stages, which is the reason why we decided to separate dwarfs from giants in our experiment.

\subsection{Cluster homogeneity}

To evaluate the viability of the chemical tagging technique, we performed a blind chemical tagging experiment designed to recover the initial stellar groups by only using the chemical abundances and the total number of clusters. We note that we discarded stars considered outliers based on their chemical signature, which should make the task easier than a real scenario with unknown stellar groups (i.e., field stars). 

The K-Means method \citep[also known as Lloyd's algorithm,][]{1056489} is a well-established machine learning/clustering algorithm for which the benefits and drawbacks have already been widely studied. Its simplicity and the use of the number of clusters that we want to find as an input parameter makes this algorithm ideal for our experiment. 

K-Means aims to partition the observations (one observation would be one star with its measured abundances) into K clusters in which each observation belongs to the cluster with the nearest mean (called centroid). Given a fixed number of clusters, K-Means clustering is reduced to an optimization problem where it finds the K centroids and assigns the observations to the nearest one, such that the squared distances are minimized.

The election of the number of clusters is usually a limitation if no a priori information is available, but in our case we provided the real number of clusters. Other known drawbacks are that this algorithm tends to find clusters of comparable spatial extent, it often incorrectly cuts the borders in between clusters (the algorithm optimizes cluster centers, not cluster borders), and the final results might depend on the initial position of the centroids. To address the last drawback, we used a variation of the algorithm named K-Means++ which optimizes the position of the initially random centroid with a probability proportional to its squared distance from the closest observation \citep{arthur2007k}.

To evaluate the goodness of the clustering, we used the following well-known metrics

\begin{enumerate}
    \item V-measure\footnote{The 'V' stands for "validity", in the sense of the goodness of a clustering solution.} \citep{hirschberg2007v} is a harmonic mean between homogeneity and completeness:
        \begin{equation}
             \mathrm{v} = 2 \times \frac{\mathrm{homogeneity} \times \mathrm{completeness}}{\mathrm{homogeneity} + \mathrm{completeness}},
        \end{equation}

        \begin{enumerate}
            \item Homogeneity: A clustering result satisfies homogeneity if all of its predicted clusters contain only data points that are members of one real open cluster. Score is between 0.0 and 1.0 and the latter stands for perfectly homogeneous labeling.

            \item Completeness: A clustering result satisfies completeness if all the data points that are members of a given real open cluster are elements of the same predicted cluster. Score is between 0.0 and 1.0 and the latter stands for perfectly complete labeling.
        \end{enumerate}

    \item Silhouette coefficient \citep{rousseeuw1987silhouettes}: Measures the concepts of cluster cohesion (favoring models that contain tightly cohesive clusters) and cluster separation (favoring models that contain highly separated clusters). The coefficient is calculated as

        \begin{equation}
            s = \frac{d_{\mathrm{inter}} - d_{\mathrm{intra}}}{\mathrm{max}\left(d_{\mathrm{intra}}, d_{\mathrm{inter}}\right)},
        \end{equation}

        where $d_{\mathrm{intra}}$ is the mean intra-cluster distance and $d_{\mathrm{inter}}$ is the mean nearest-cluster distance. Negative values (never smaller than -1) indicate that most of the stars are assigned to a incorrect cluster, values close to zero indicate the existence of overlapping clusters, and values close to 1.0 indicate that the stars of a given cluster are similar to each other and well-separated from other clusters.
\end{enumerate}

The clustering configurations that maximizes the V-measure and the mean silhouette coefficient is the best although it should be taken into account that the silhouette is usually reduced when adding more dimensions. We found the best results when the clustering algorithm is run separately in groups divided by stellar types (i.e., dwarfs and giants) and with the first five principal components (built from the 17 abundances; see Table \ref{tab:clustering_metrics}).

It is interesting to see that we cover 85\% of the variance with four or five principal components, contrary to \cite{2012MNRAS.421.1231T} where six or seven components are needed. 
Our abundances were derived homogeneously, but our sample is smaller.
We have 31 clusters and \cite{2012MNRAS.421.1231T} analyze 78 clusters, which can lead to a    
bigger diversity and thus, more components are needed to cover the same          
amount of variance.                        

For all the configurations, the silhouette coefficient is under 0.50, indicating that the structure found is reasonable but weak. There is a non-negligible chemical overlapping among stars of different clusters.

\begin{table}[ht!]
    \begin{center}
        \caption{K-Means++ clustering metrics for different stellar groups and dimensions.}
        \label{tab:clustering_metrics}
        \tabcolsep=0.08cm
        \begin{tabular}{l c | c | c | c c | c c | c c | c c}

    &       &    & \scriptsize{Var.}   &   \multicolumn{2}{c|}{\scriptsize{Homogeneity}}            &   \multicolumn{2}{c|}{\scriptsize{Completeness}}            &   \multicolumn{2}{c|}{\scriptsize{V-Measure}}            &   \multicolumn{2}{c}{\scriptsize{Silhouette}}            \\
    &       &   D   &  &  $\bar{x}$   &   $\sigma$    &   $\bar{x}$   &   $\sigma$    &   $\bar{x}$   &   $\sigma$    &   $\bar{x}$   &   $\sigma$    \\

\parbox[t]{2mm}{\multirow{18}{*}{\rotatebox[origin=c]{90}{All}}}    &   \parbox[t]{2mm}{\multirow{18}{*}{\rotatebox[origin=c]{90}{\scriptsize{31 clusters (177 stars)}}}}   &   2   &   72.9\%   &   0.72    &   0.01    &   0.61    &   0.01    &   0.66    &   0.01    &   0.40    &   0.01    \\
    &       &   3   &   82.2\%   &   0.79    &   0.01    &   0.67    &   0.01    &   0.73    &   0.01    &   0.36    &   0.01    \\
    &       &   4   &   86.1\%   &   0.79    &   0.01    &   0.68    &   0.01    &   0.73    &   0.01    &   0.32    &   0.01    \\
    &       &   5   &   88.8\%   &   0.78    &   0.01    &   0.68    &   0.01    &   0.72    &   0.01    &   0.29    &   0.01    \\
    &       &   6   &   91.0\%   &   0.78    &   0.01    &   0.68    &   0.01    &   0.73    &   0.01    &   0.27    &   0.01    \\
    &       &   7   &   93.1\%   &   0.78    &   0.01    &   0.69    &   0.01    &   0.73    &   0.01    &   0.26    &   0.01    \\
    &       &   8   &   94.6\%   &   0.78    &   0.01    &   0.69    &   0.01    &   0.73    &   0.01    &   0.24    &   0.01    \\
    &       &   9   &   95.6\%   &   0.77    &   0.01    &   0.68    &   0.01    &   0.73    &   0.01    &   0.24    &   0.01    \\
    &       &   10  &   96.6\%   &   0.77    &   0.01    &   0.68    &   0.01    &   0.73    &   0.01    &   0.23    &   0.01    \\
    &       &   11  &   97.5\%   &   0.77    &   0.01    &   0.68    &   0.01    &   0.72    &   0.01    &   0.23    &   0.01    \\
    &       &   12  &   98.2\%   &   0.77    &   0.01    &   0.68    &   0.01    &   0.72    &   0.01    &   0.22    &   0.01    \\
    &       &   13  &   98.8\%   &   0.77    &   0.01    &   0.68    &   0.01    &   0.72    &   0.01    &   0.21    &   0.01    \\
    &       &   14  &   99.2\%   &   0.77    &   0.01    &   0.69    &   0.01    &   0.72    &   0.01    &   0.21    &   0.01    \\
    &       &   15  &   99.6\%   &   0.77    &   0.01    &   0.69    &   0.01    &   0.72    &   0.01    &   0.21    &   0.01    \\
    &       &   16  &   99.9\%   &   0.77    &   0.01    &   0.69    &   0.01    &   0.73    &   0.01    &   0.21    &   0.01    \\
    &       &   17  &   100.0\%  &   0.76    &   0.02    &   0.68    &   0.01    &   0.72    &   0.01    &   0.20    &   0.01    \\
    &       &   17$^*$  &   100.0\%  &   0.76    &   0.02    &   0.68    &   0.01    &   0.72    &   0.01    &   0.20    &   0.01    \\
\hline                                                                                          
\parbox[t]{2mm}{\multirow{18}{*}{\rotatebox[origin=c]{90}{Dwarfs}}} &   \parbox[t]{2mm}{\multirow{18}{*}{\rotatebox[origin=c]{90}{\scriptsize{11 clusters (60 stars)}}}}    &   2   &   62.9\%   &   0.81    &   0.01    &   0.61    &   0.01    &   0.69    &   0.01    &   0.42    &   0.01    \\
    &       &   3   &   73.3\%   &   0.75    &   0.01    &   0.57    &   0.01    &   0.65    &   0.01    &   0.38    &   0.01    \\
    &       &   4   &   79.9\%   &   0.83    &   0.02    &   0.63    &   0.02    &   0.71    &   0.02    &   0.34    &   0.01    \\
    &       &   5   &   84.3\%   &   0.83    &   0.02    &   0.63    &   0.02    &   0.72    &   0.02    &   0.32    &   0.01    \\
    &       &   6   &   88.1\%   &   0.78    &   0.03    &   0.59    &   0.03    &   0.67    &   0.03    &   0.28    &   0.01    \\
    &       &   7   &   91.1\%   &   0.78    &   0.03    &   0.60    &   0.02    &   0.68    &   0.03    &   0.24    &   0.01    \\
    &       &   8   &   93.6\%   &   0.79    &   0.04    &   0.61    &   0.03    &   0.69    &   0.03    &   0.23    &   0.01    \\
    &       &   9   &   95.4\%   &   0.79    &   0.04    &   0.61    &   0.03    &   0.69    &   0.03    &   0.21    &   0.01    \\
    &       &   10  &   96.7\%   &   0.78    &   0.04    &   0.61    &   0.03    &   0.68    &   0.03    &   0.21    &   0.01    \\
    &       &   11  &   97.5\%   &   0.77    &   0.04    &   0.60    &   0.03    &   0.67    &   0.03    &   0.20    &   0.01    \\
    &       &   12  &   98.2\%   &   0.77    &   0.04    &   0.60    &   0.03    &   0.68    &   0.03    &   0.20    &   0.01    \\
    &       &   13  &   98.8\%   &   0.77    &   0.04    &   0.60    &   0.03    &   0.68    &   0.04    &   0.19    &   0.01    \\
    &       &   14  &   99.2\%   &   0.77    &   0.04    &   0.60    &   0.03    &   0.67    &   0.03    &   0.19    &   0.01    \\
    &       &   15  &   99.6\%   &   0.78    &   0.04    &   0.61    &   0.03    &   0.68    &   0.04    &   0.19    &   0.01    \\
    &       &   16  &   99.9\%   &   0.77    &   0.04    &   0.60    &   0.03    &   0.67    &   0.03    &   0.19    &   0.01    \\
    &       &   17  &   100.0\%  &   0.77    &   0.04    &   0.60    &   0.03    &   0.67    &   0.03    &   0.18    &   0.01    \\
    &       &   17$^*$  &   100.0\%  &   0.77    &   0.04    &   0.60    &   0.03    &   0.67    &   0.03    &   0.18    &   0.01    \\
\hline                                                                                          
\parbox[t]{2mm}{\multirow{18}{*}{\rotatebox[origin=c]{90}{Giants}}} &   \parbox[t]{2mm}{\multirow{18}{*}{\rotatebox[origin=c]{90}{\scriptsize{26 clusters (117 stars)}}}}   &   2   &   71.1\%   &   0.71    &   0.01    &   0.64    &   0.01    &   0.67    &   0.01    &   0.35    &   0.01    \\
    &       &   3   &   83.4\%   &   0.79    &   0.01    &   0.71    &   0.01    &   0.75    &   0.01    &   0.33    &   0.01    \\
    &       &   4   &   87.7\%   &   0.79    &   0.01    &   0.72    &   0.01    &   0.75    &   0.01    &   0.29    &   0.01    \\
    &       &   5   &   90.4\%   &   0.81    &   0.01    &   0.74    &   0.01    &   0.78    &   0.01    &   0.28    &   0.02    \\
    &       &   6   &   92.6\%   &   0.80    &   0.01    &   0.73    &   0.01    &   0.76    &   0.01    &   0.26    &   0.01    \\
    &       &   7   &   94.1\%   &   0.80    &   0.01    &   0.74    &   0.01    &   0.77    &   0.01    &   0.24    &   0.01    \\
    &       &   8   &   95.4\%   &   0.81    &   0.01    &   0.74    &   0.01    &   0.77    &   0.01    &   0.24    &   0.01    \\
    &       &   9   &   96.4\%   &   0.81    &   0.01    &   0.74    &   0.01    &   0.77    &   0.01    &   0.23    &   0.01    \\
    &       &   10  &   97.2\%   &   0.81    &   0.01    &   0.75    &   0.01    &   0.78    &   0.01    &   0.22    &   0.01    \\
    &       &   11  &   98.0\%   &   0.81    &   0.01    &   0.75    &   0.02    &   0.78    &   0.01    &   0.22    &   0.01    \\
    &       &   12  &   98.6\%   &   0.81    &   0.01    &   0.76    &   0.01    &   0.78    &   0.01    &   0.22    &   0.01    \\
    &       &   13  &   99.1\%   &   0.81    &   0.02    &   0.76    &   0.01    &   0.78    &   0.01    &   0.21    &   0.01    \\
    &       &   14  &   99.5\%   &   0.81    &   0.01    &   0.75    &   0.01    &   0.78    &   0.01    &   0.21    &   0.01    \\
    &       &   15  &   99.8\%   &   0.81    &   0.01    &   0.75    &   0.02    &   0.78    &   0.01    &   0.21    &   0.01    \\
    &       &   16  &   99.9\%   &   0.81    &   0.01    &   0.75    &   0.01    &   0.78    &   0.01    &   0.20    &   0.01    \\
    &       &   17  &   100.0\%  &   0.81    &   0.01    &   0.76    &   0.01    &   0.78    &   0.01    &   0.21    &   0.01    \\
    &       &   17$^*$  &   100.0\%  &   0.81    &   0.01    &   0.76    &   0.01    &   0.78    &   0.01    &   0.21    &   0.01    \\

        \end{tabular}
    \end{center}
    \tablefoot{ The number of dimension (D) refers to the number of principal components used, except those marked with $^*$ where the non-transformed abundances were used. The covered variance (Var.) is presented in the second column. The V-Measure is calculated from the homogeneity and completeness values. All the parameters have an average value and a standard deviation that comes from 100 random iterations.}
\end{table}

\begin{table}[ht!]
    \begin{center}
        \caption{Benchmark of different clustering algorithms separated by stellar group.}
        \label{tab:clustering_metrics_benchmark}
        \tabcolsep=0.10cm
        \begin{tabular}{l l | l | c | c | c c | c | c | c }

    &       &               &   D   &   N   &   H           &   C           &   V           &   S           &   $\star$         \\
\parbox[t]{2mm}{\multirow{8}{*}{\rotatebox[origin=c]{90}{All}}} &   \parbox[t]{2mm}{\multirow{8}{*}{\rotatebox[origin=c]{90}{\scriptsize{31 clusters (177 stars)}}}}  &   K-Means++           &   5   &   -   &   0.78            &   0.68            &   0.72            &   0.29            &   100\%            \\
    &       &   Affinity prop.            &   5   &   29  &   0.77            &   0.69            &   0.72            &   0.30            &   100\%            \\
    &       &   DBSCAN          &   5   &   30  &   0.48            &   0.70            &   0.57            &   0.03            &   100\%            \\
    &       &   Mitschang \scriptsize{(0.68 / 0)}            &   17$^*$  &   42  &   0.85            &   0.66            &   0.74            &   0.11            &   100\%            \\
    &       &   Mitschang \scriptsize{(0.68 / 2)}            &   17$^*$  &   32  &   0.84            &   0.66            &   0.74            &   0.12            &   89\%         \\
    &       &   Mitschang \scriptsize{(0.90 / 0)}            &   17$^*$  &   42  &   0.86            &   0.66            &   0.75            &   0.11            &   97\%         \\
    &       &   Mitschang \scriptsize{(0.90 / 2)}            &   17$^*$  &   32  &   0.84            &   0.66            &   0.74            &   0.13            &   86\%         \\
\hline                                                                                                                                                                                                      
\parbox[t]{2mm}{\multirow{8}{*}{\rotatebox[origin=c]{90}{Dwarfs}}}  &   \parbox[t]{2mm}{\multirow{8}{*}{\rotatebox[origin=c]{90}{\scriptsize{11 clusters (60 stars)}}}}   &   K-Means++          &   5   &   -   &   0.83            &   0.63            &   0.72            &   0.32            &   100\%            \\
    &       &   Affinity prop.            &   5   &   13  &   0.85            &   0.61            &   0.71            &   0.31            &   100\%            \\
    &       &   DBSCAN          &   5   &   13  &   0.59            &   0.69            &   0.64            &   0.01            &   100\%            \\
    &       &   Mitschang \scriptsize{(0.68 / 0)}            &   17$^*$  &   11  &   0.82            &   0.59            &   0.69            &   0.07            &   98\%         \\
    &       &   Mitschang \scriptsize{(0.68 / 2)}            &   17$^*$  &   10  &   0.81            &   0.59            &   0.68            &   0.11            &   95\%         \\
    &       &   Mitschang \scriptsize{(0.90 / 0)}            &   17$^*$  &   11  &   0.84            &   0.59            &   0.69            &   0.09            &   90\%         \\
    &       &   Mitschang \scriptsize{(0.90 / 2)}            &   17$^*$  &   10  &   0.83            &   0.58            &   0.68            &   0.13            &   87\%         \\
\hline                                                                                                                                                                                                      
\parbox[t]{2mm}{\multirow{8}{*}{\rotatebox[origin=c]{90}{Giants}}}  &   \parbox[t]{2mm}{\multirow{8}{*}{\rotatebox[origin=c]{90}{\scriptsize{26 clusters (117 stars)}}}}   &   K-Means++         &   5   &   31  &   0.81            &   0.74            &   0.78            &   0.28            &   100\%            \\
    &       &   Affinity prop.            &   5   &   24  &   0.79            &   0.74            &   0.76            &   0.28            &   100\%            \\
    &       &   DBSCAN          &   5   &   28  &   0.66            &   0.83            &   0.74            &   0.06            &   100\%            \\
    &       &   Mitschang \scriptsize{(0.68 / 0)}            &   17$^*$  &   31  &   0.84            &   0.70            &   0.77            &   0.12            &   100\%            \\
    &       &   Mitschang \scriptsize{(0.68 / 2)}            &   17$^*$  &   21  &   0.82            &   0.71            &   0.76            &   0.14            &   83\%         \\
    &       &   Mitschang \scriptsize{(0.90 / 0)}            &   17$^*$  &   31  &   0.84            &   0.70            &   0.77            &   0.12            &   100\%            \\
    &       &   Mitschang \scriptsize{(0.90 / 2)}            &   17$^*$  &   21  &   0.82            &   0.71            &   0.76            &   0.14            &   83\%         \\

        \end{tabular}
    \end{center}
    \tablefoot{ The number of dimension (D) refers to the number of principal components used, except those marked with $^*$ where the non-transformed abundances were used.
The number of predicted clusters (N) is not reported for the K-Means++ algorithms because it is one of the parameters of the method and the real number was used.
 The V-Measure is calculated from the homogeneity (H) and completeness (C) values. The silhouette coefficient (S) and the number of classified stars ($\star$) are reported in the last two columns.}
\end{table}

\begin{figure}
    \begin{centering}
        \includegraphics[width=\linewidth, trim = 1mm 1mm 1mm 1mm, clip]{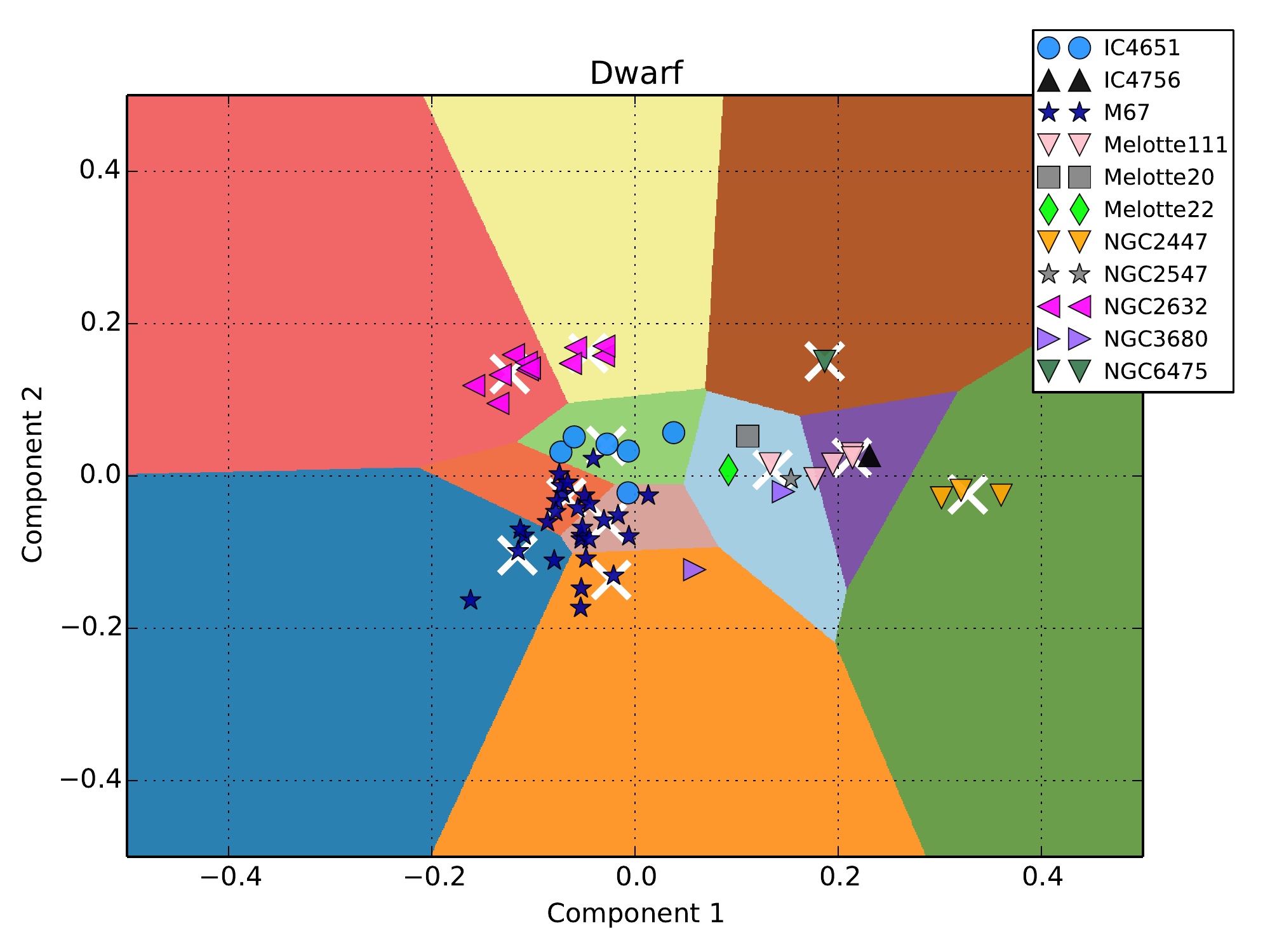}
        \par
    \end{centering}
    \caption{Dwarfs represented using the first two components of PCA. Background colors correspond to the clusters found by the K-Means algorithm. Centroids are marked with white crosses.}
    \label{fig:clustering_dwarfs}
\end{figure}

\begin{figure}
    \begin{centering}
        \includegraphics[width=\linewidth, trim = 1mm 1mm 1mm 1mm, clip]{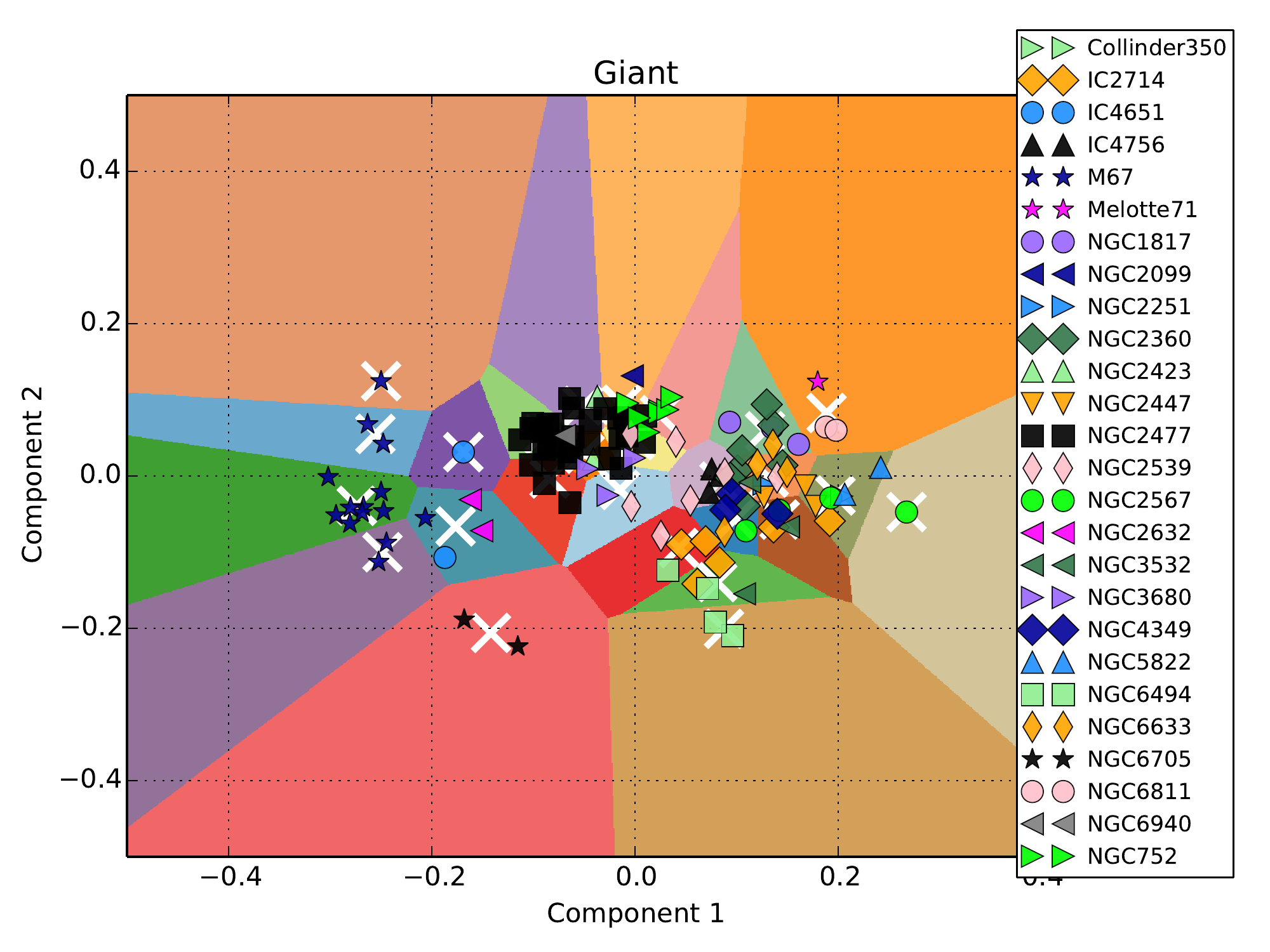}
        \par
    \end{centering}
    \caption{Giants represented using the first two components of PCA. Background colors correspond to the clusters found by the K-Means algorithm. Centroids are marked with white crosses.}
    \label{fig:clustering_giants}
\end{figure}

\begin{figure}
    \begin{centering}
        \includegraphics[width=9cm, trim = 1mm 1mm 1mm 1mm, clip]{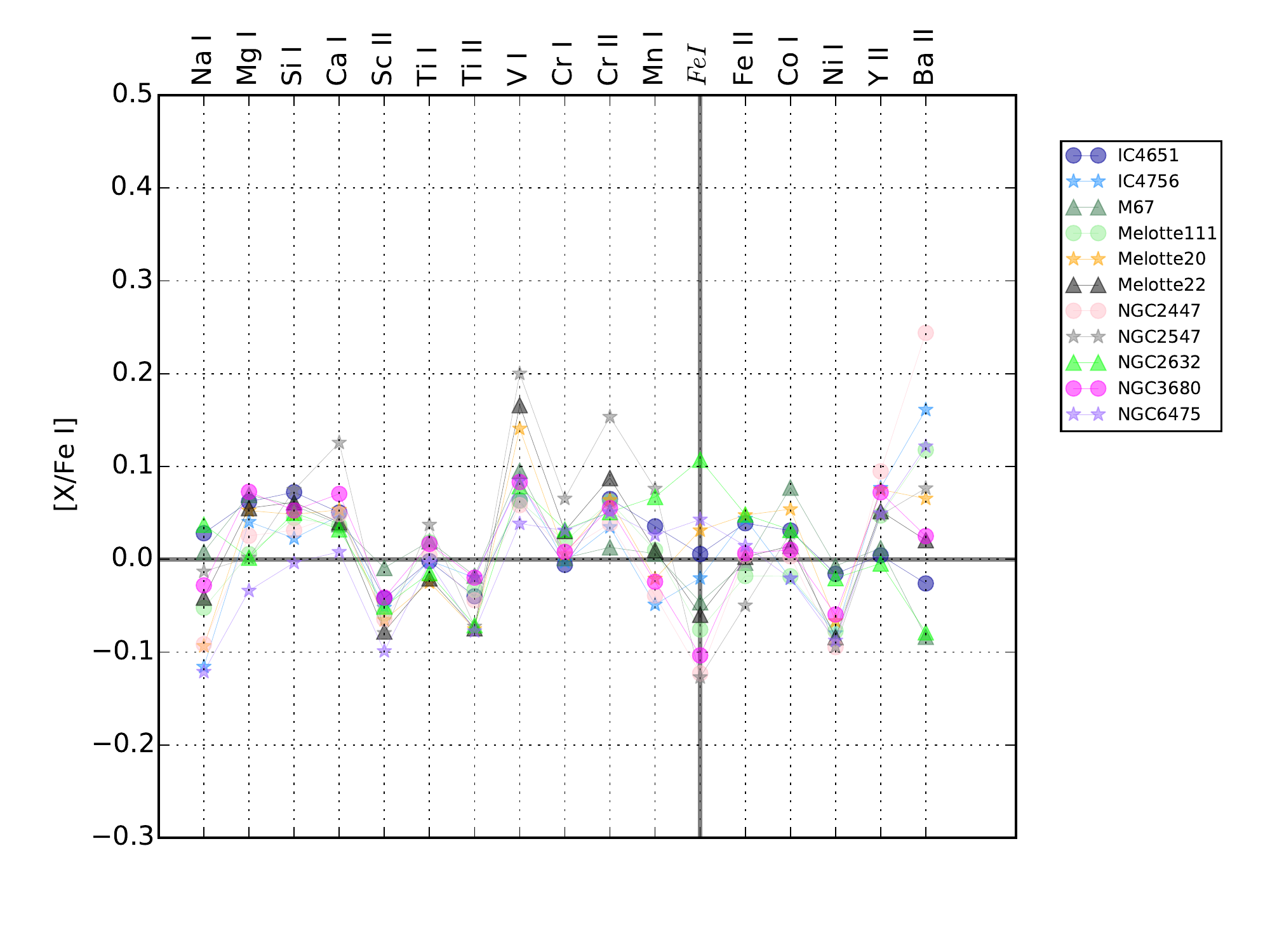}
    \end{centering}
    \caption{Stellar abundances averaged per cluster using only dwarf stars.}
    \label{fig:abundance_overlapping1}
\end{figure}

\begin{figure}
    \begin{centering}
        \includegraphics[width=9cm, trim = 1mm 1mm 1mm 1mm, clip]{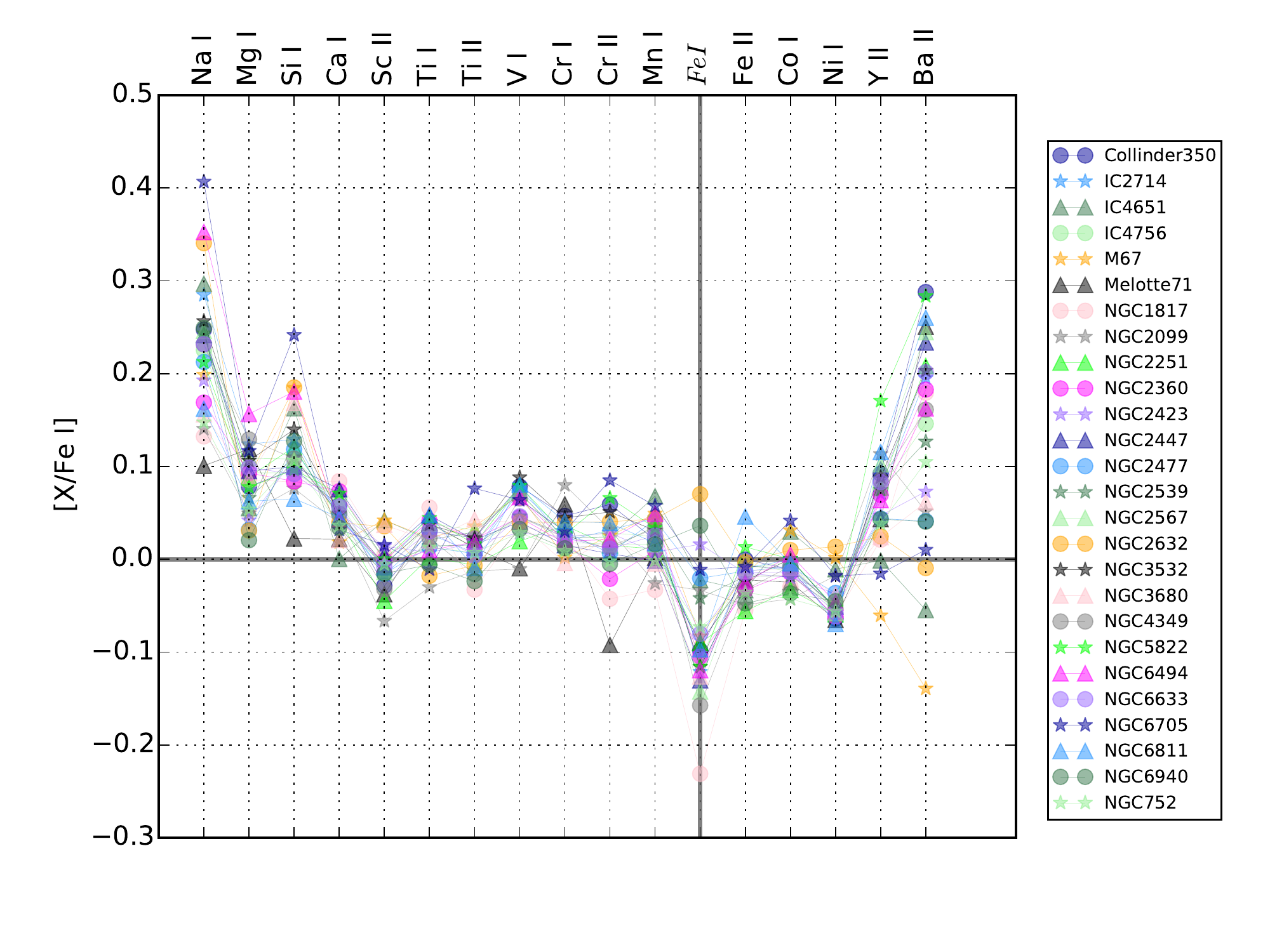}
    \end{centering}
    \caption{Stellar abundances averaged per cluster using only giant stars.}
    \label{fig:abundance_overlapping2}
\end{figure}

The clustering analysis groups together stars from different open clusters (see Tables \ref{tab:clustering_dwarfs} and \ref{tab:clustering_giants}; Figs. \ref{fig:clustering_dwarfs} and \ref{fig:clustering_giants}) pointing out that, for the analyzed elemental abundances, the clusters' chemical signature are not significantly different. The abundance patterns change relatively smoothly in the chemical space as shown in Figs. \ref{fig:abundance_overlapping1} and \ref{fig:abundance_overlapping2} (especially considering that the typical uncertainty is $\sim$0.05 dex), complicating the separation of stars that belong to different clusters.

In addition to the K-Means method, we tested other known machine learning algorithms such as affinity propagation \citep{frey2007clustering} and DBSCAN \citep{ester1996density}. These do not require specifying the number of clusters to be found as an input, but need other distance parameters that are going to have a significant impact on the number of clusters automatically found by the algorithm. In our case, to do a fair comparison with K-Means we have fine-tuned the input parameters to obtain a number of clusters close to the real one. As input values, we used the first five principal components, which cover approximately 85\% of the variance. To complement the comparison, we also implemented the classification algorithm developed by \cite{2013MNRAS.428.2321M} which does not need any extra input parameters and it also automatically finds the number of clusters. For this method we use the 17 elements directly because this approach was calibrated with real abundances and not principal components. Four different configurations have been executed using the following criteria: limiting the probability of belonging to the cluster to be higher than 68\% or 90\%; discarding clusters found with only two stars or not discarding any resulting cluster. The results are shown in Table \ref{tab:clustering_metrics_benchmark}; all the methods have a similar behavior in terms of V-Measure except DBSCAN, which is the worse behaving with dwarf stars (it might be more sensitive to the lower number of stars).

To estimate an order of magnitude of the star contamination level, we linked each group (found by the clustering algorithms) to the open cluster that contains the highest number of stars (unless the open cluster was already assigned) and we found that around 30-50\% of the stars are not assigned to their expected cluster. The higher the number of open clusters included in the analysis, the higher the contamination percentage. Considering that it has been estimated that about 10$^{8}$ clusters were dissolved in the Galaxy \citep{2004PASA...21..110B}, we expect a severe contamination when applying the chemical tagging technique to field stars in order to recover co-natal aggregates.

\begin{figure}
    \begin{centering}
        \includegraphics[width=9cm, trim = 1mm 1mm 1mm 1mm, clip]{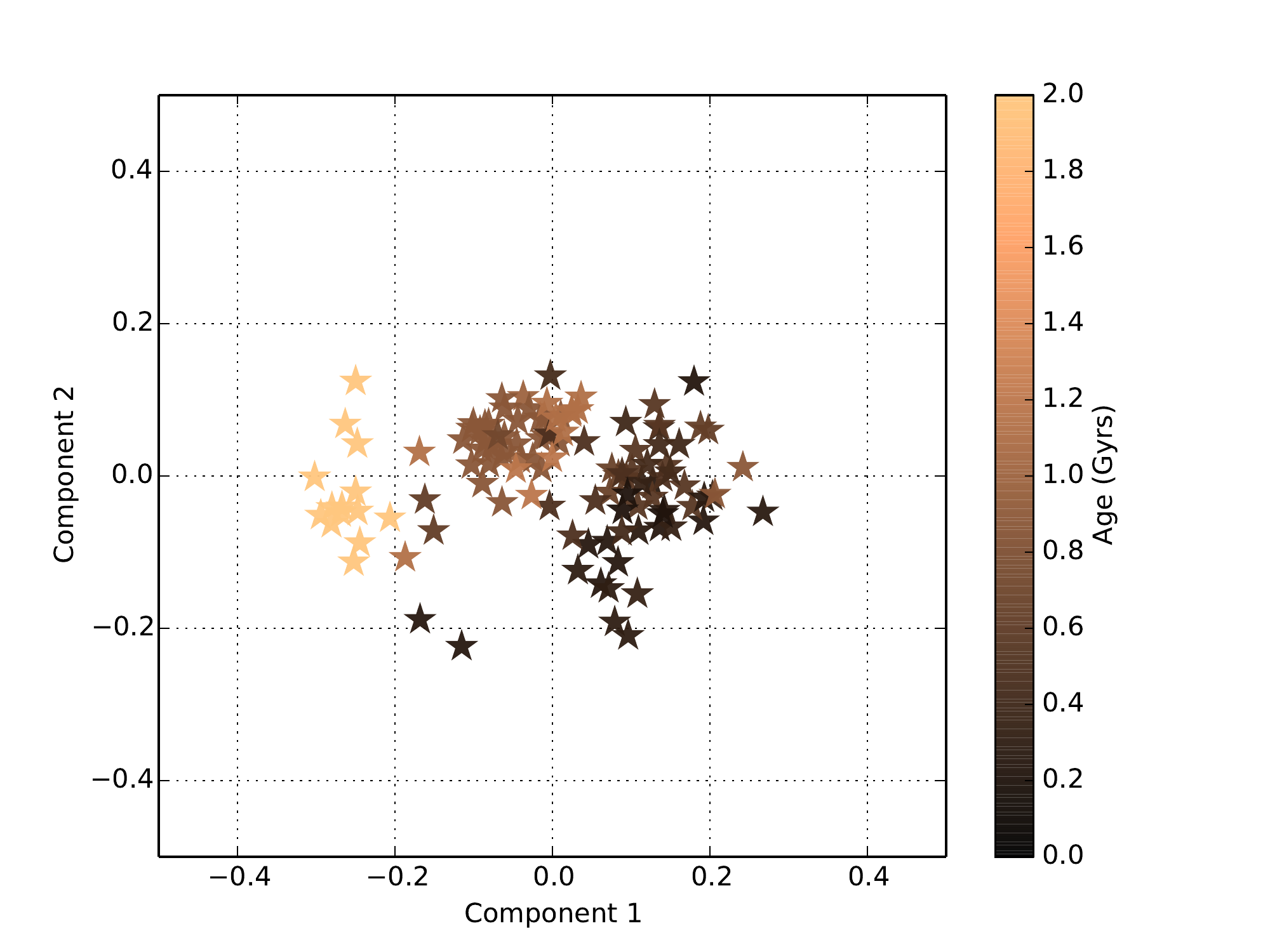}
    \end{centering}
    \caption{Stars represented in the PCA space using the first two principal components. The stellar ages are indicated by the color scale bar.}
    \label{fig:pca_age}
\end{figure}

For the giant case, where the number of stars is higher, there might be a correlation between the first two principal components and the stellar ages (i.e., cluster ages). In Fig. \ref{fig:pca_age} we see that the stars with a lower value in the first two components belong to younger clusters, while older stars have lower values in the first component but higher in the second one. The correlation seems to follow a semi-circular path in that visualization of the PCA space (although the correlation seems stronger for the first component). Thus, this keeps open the possibility of finding co-eval aggregates (stars born at the same period although not necessarily from the same molecular cloud) by using the chemical tagging technique.

\begin{table}[ht!]
    \begin{center}
        \caption{Clusters found for dwarf stars and their real open clusters using five components.}
        \label{tab:clustering_dwarfs}
        \begin{tabular}{c|c |c| c c | c }
        &   &   &   \multicolumn{2}{c|}{Coverage}  &  Age  \\
        Cluster   &   OC &   Stars  &   Cluster  &   OC &   (Gyrs) \\

        \hline

\multirow{  1   }{*}{N. 01  D   }   &   IC4651  &   6   &   100\%    &   100\%    &   1.14    \\
                            \hline                                  
\multirow{  2   }{*}{N. 02  D   }   &   M67 &   8   &   89\% &   29\% &   4.09    \\
                        &   NGC3680 &   1   &   11\% &   50\% &   1.19    \\
                            \hline                                  
\multirow{  4   }{*}{N. 03  D   }   &   Melotte111  &   4   &   57\% &   80\% &   0.71    \\
                        &   IC4756  &   1   &   14\% &   100\%    &   0.67    \\
                        &   Melotte20   &   1   &   14\% &   100\%    &   0.06    \\
                        &   NGC6475 &   1   &   14\% &   100\%    &   0.25    \\
                            \hline                                  
\multirow{  4   }{*}{N. 04  D   }   &   Melotte111  &   1   &   25\% &   20\% &   0.71    \\
                        &   Melotte22   &   1   &   25\% &   100\%    &   0.11    \\
                        &   NGC2547 &   1   &   25\% &   100\%    &   0.04    \\
                        &   NGC3680 &   1   &   25\% &   50\% &   1.19    \\
                            \hline                                  
\multirow{  1   }{*}{N. 05  D   }   &   M67 &   12  &   100\%    &   43\% &   4.09    \\
                            \hline                                  
\multirow{  1   }{*}{N. 06  D   }   &   M67 &   4   &   100\%    &   14\% &   4.09    \\
                            \hline                                  
\multirow{  1   }{*}{N. 07  D   }   &   M67 &   2   &   100\%    &   7\%  &   4.09    \\
                            \hline                                  
\multirow{  1   }{*}{N. 08  D   }   &   M67 &   2   &   100\%    &   7\%  &   4.09    \\
                            \hline                                  
\multirow{  1   }{*}{N. 09  D   }   &   NGC2447 &   3   &   100\%    &   100\%    &   0.56    \\
                            \hline                                  
\multirow{  1   }{*}{N. 10  D   }   &   NGC2632 &   7   &   100\%    &   64\% &   0.63    \\
                            \hline                                  
\multirow{  1   }{*}{N. 11  D   }   &   NGC2632 &   4   &   100\%    &   36\% &   0.63    \\

        \end{tabular}
    \end{center}
    \tablefoot{ The clustering algorithm assigns numbers to each identified group as indicated in the first column. The number of stars are presented next to the real open cluster to which they belong. The coverage indicates the percentage of stars found by comparing the total number of stars in the identified group to the number in the real open cluster. }
\end{table}

\begin{table}[ht!]
    \begin{center}
        \caption{Clusters found for giant stars and their real open clusters using five components.}
        \label{tab:clustering_giants}
        \small
        \begin{tabular}{c|c |c| c c | c }
        &   &   &   \multicolumn{2}{c|}{Coverage}  &  Age  \\
        Cluster   &   OC &   Stars  &   Cluster  &   OC &   (Gyrs) \\
        \hline

\multirow{  4   }{*}{N. 01  G   }   &   IC2714  &   2   &   29\% &   33\% &   0.25    \\
                        &   NGC2447 &   2   &   29\% &   67\% &   0.56    \\
                        &   NGC2567 &   2   &   29\% &   50\% &   0.28    \\
                        &   NGC2539 &   1   &   14\% &   14\% &   0.49    \\
                            \hline                                  
\multirow{  2   }{*}{N. 02  G   }   &   NGC2477 &   10  &   91\% &   34\% &   0.88    \\
                        &   IC4651  &   1   &   9\%  &   50\% &   1.14    \\
                            \hline                                  
\multirow{  2   }{*}{N. 03  G   }   &   IC2714  &   4   &   80\% &   67\% &   0.25    \\
                        &   NGC6494 &   1   &   20\% &   25\% &   0.30    \\
                            \hline                                  
\multirow{  1   }{*}{N. 04  G   }   &   NGC752  &   7   &   100\%    &   100\%    &   1.12    \\
                            \hline                                  
\multirow{  3   }{*}{N. 05  G   }   &   Collinder350    &   1   &   33\% &   100\%    &   0.41    \\
                        &   NGC2567 &   1   &   33\% &   25\% &   0.28    \\
                        &   NGC5822 &   1   &   33\% &   50\% &   0.89    \\
                            \hline                                  
\multirow{  1   }{*}{N. 06  G   }   &   NGC6705 &   2   &   100\%    &   100\%    &   0.25    \\
                            \hline                                  
\multirow{  1   }{*}{N. 07  G   }   &   NGC1817 &   3   &   100\%    &   100\%    &   0.41    \\
                            \hline                                  
\multirow{  2   }{*}{N. 08  G   }   &   NGC3532 &   1   &   50\% &   33\% &   0.35    \\
                        &   NGC5822 &   1   &   50\% &   50\% &   0.89    \\
                            \hline                                  
\multirow{  2   }{*}{N. 09  G   }   &   NGC2539 &   2   &   67\% &   29\% &   0.49    \\
                        &   NGC6633 &   1   &   33\% &   25\% &   0.43    \\
                            \hline                                  
\multirow{  4   }{*}{N. 10  G   }   &   NGC6633 &   3   &   50\% &   75\% &   0.43    \\
                        &   NGC2251 &   1   &   17\% &   100\%    &   0.27    \\
                        &   NGC2447 &   1   &   17\% &   33\% &   0.56    \\
                        &   NGC3532 &   1   &   17\% &   33\% &   0.35    \\
                            \hline                                  
\multirow{  2   }{*}{N. 11  G   }   &   NGC2360 &   2   &   67\% &   29\% &   0.56    \\
                        &   Melotte71   &   1   &   33\% &   100\%    &   0.23    \\
                            \hline                                  
\multirow{  3   }{*}{N. 12  G   }   &   NGC4349 &   3   &   60\% &   100\%    &   0.21    \\
                        &   NGC2360 &   1   &   20\% &   14\% &   0.56    \\
                        &   NGC2567 &   1   &   20\% &   25\% &   0.28    \\
                            \hline                                  
\multirow{  1   }{*}{N. 13  G   }   &   M67 &   10  &   100\%    &   71\% &   4.09    \\
                            \hline                                  
\multirow{  1   }{*}{N. 14  G   }   &   M67 &   3   &   100\%    &   21\% &   4.09    \\
                            \hline                                  
\multirow{  1   }{*}{N. 15  G   }   &   M67 &   1   &   100\%    &   7\%  &   4.09    \\
                            \hline                                  
\multirow{  1   }{*}{N. 16  G   }   &   NGC3680 &   1   &   100\%    &   33\% &   1.19    \\
                            \hline                                  
\multirow{  1   }{*}{N. 17  G   }   &   NGC3680 &   2   &   100\%    &   67\% &   1.19    \\
                            \hline                                  
\multirow{  3   }{*}{N. 18  G   }   &   NGC2477 &   8   &   80\% &   28\% &   0.88    \\
                        &   NGC2423 &   1   &   10\% &   50\% &   1.00    \\
                        &   NGC6940 &   1   &   10\% &   100\%    &   0.72    \\
                            \hline                                  
\multirow{  2   }{*}{N. 19  G   }   &   NGC6494 &   3   &   75\% &   75\% &   0.30    \\
                        &   NGC3532 &   1   &   25\% &   33\% &   0.35    \\
                            \hline                                  
\multirow{  1   }{*}{N. 20  G   }   &   IC4651  &   1   &   100\%    &   50\% &   1.14    \\
                            \hline                                  
\multirow{  2   }{*}{N. 21  G   }   &   NGC2477 &   7   &   88\% &   24\% &   0.88    \\
                        &   NGC2539 &   1   &   13\% &   14\% &   0.49    \\
                            \hline                                  
\multirow{  1   }{*}{N. 22  G   }   &   NGC6811 &   2   &   100\%    &   100\%    &   0.64    \\
                            \hline                                  
\multirow{  2   }{*}{N. 23  G   }   &   NGC2360 &   4   &   80\% &   57\% &   0.56    \\
                        &   IC4756  &   1   &   20\% &   33\% &   0.67    \\
                            \hline                                  
\multirow{  1   }{*}{N. 24  G   }   &   NGC2632 &   2   &   100\%    &   100\%    &   0.63    \\
                            \hline                                  
\multirow{  2   }{*}{N. 25  G   }   &   IC4756  &   2   &   50\% &   67\% &   0.67    \\
                        &   NGC2539 &   2   &   50\% &   29\% &   0.49    \\
                            \hline                                  
\multirow{  4   }{*}{N. 26  G   }   &   NGC2477 &   4   &   57\% &   14\% &   0.88    \\
                        &   NGC2099 &   1   &   14\% &   100\%    &   0.45    \\
                        &   NGC2423 &   1   &   14\% &   50\% &   1.00    \\
                        &   NGC2539 &   1   &   14\% &   14\% &   0.49    \\

        \end{tabular}
    \end{center}
    \tablefoot{ The column description is the same as for Table \ref{tab:clustering_dwarfs} }
\end{table}

\subsection{The role of different elements}

The selection of measured abundances also has an impact on the clustering algorithms, hence on the potential of the chemical tagging technique. The best elements among  those included in this study are those that can be measured with high precision (i.e., low dispersion) and show no correlation among them (e.g., alpha elements have similar trends). It is also important to have elements produced in different processes with significantly different yields.

Working with high-resolution and high S/N spectra contributes to better precision, even though not all the elements can be easily measured for all kinds of stellar types because their absorption lines can be too weak or highly blended. In this context, high-quality atomic data and reliable physical models are fundamental.

\begin{figure}
    \begin{centering}
        \includegraphics[width=\linewidth, trim = 1mm 1mm 1mm 1mm, clip]{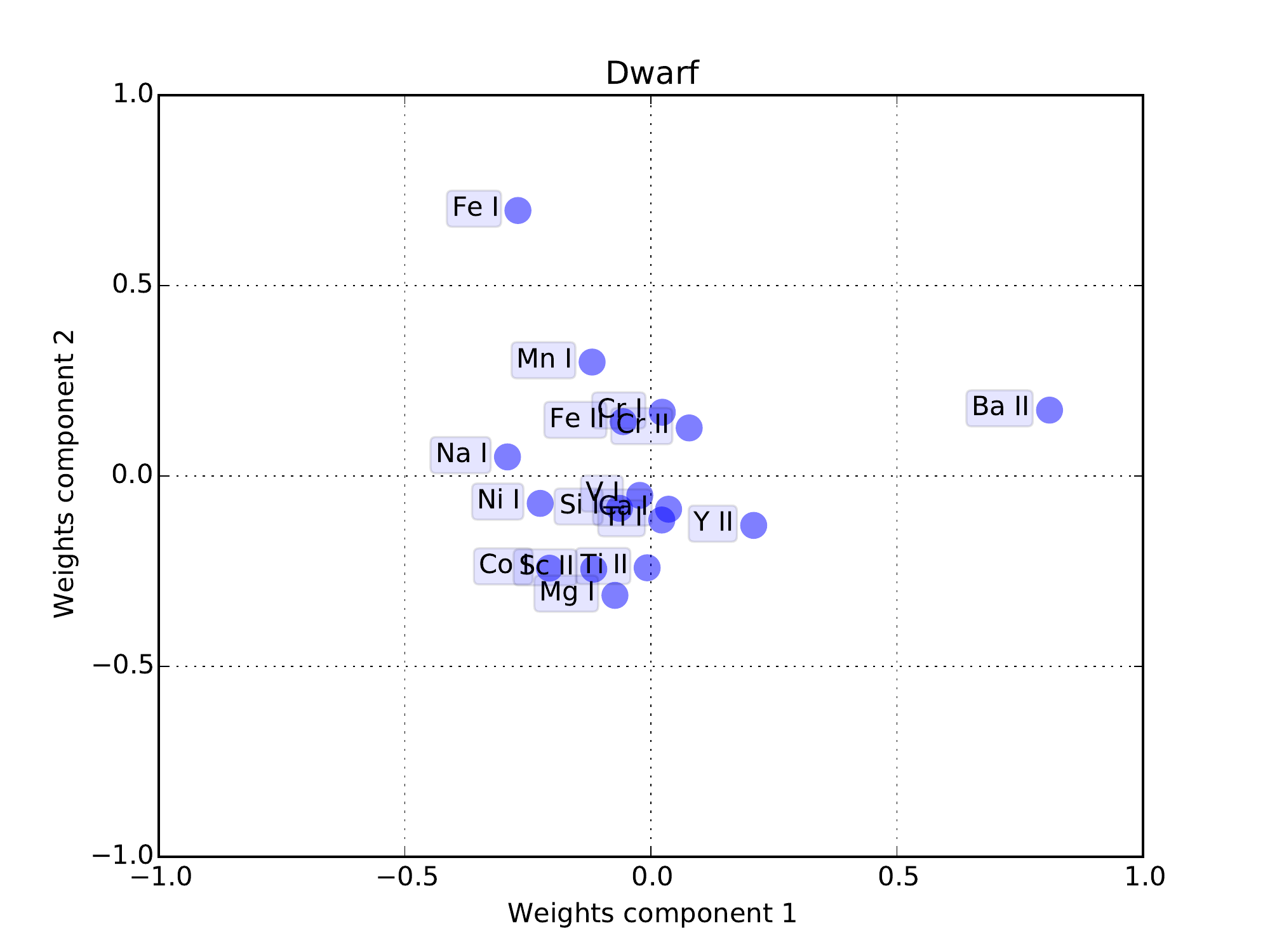}
        \includegraphics[width=\linewidth, trim = 1mm 1mm 1mm 1mm, clip]{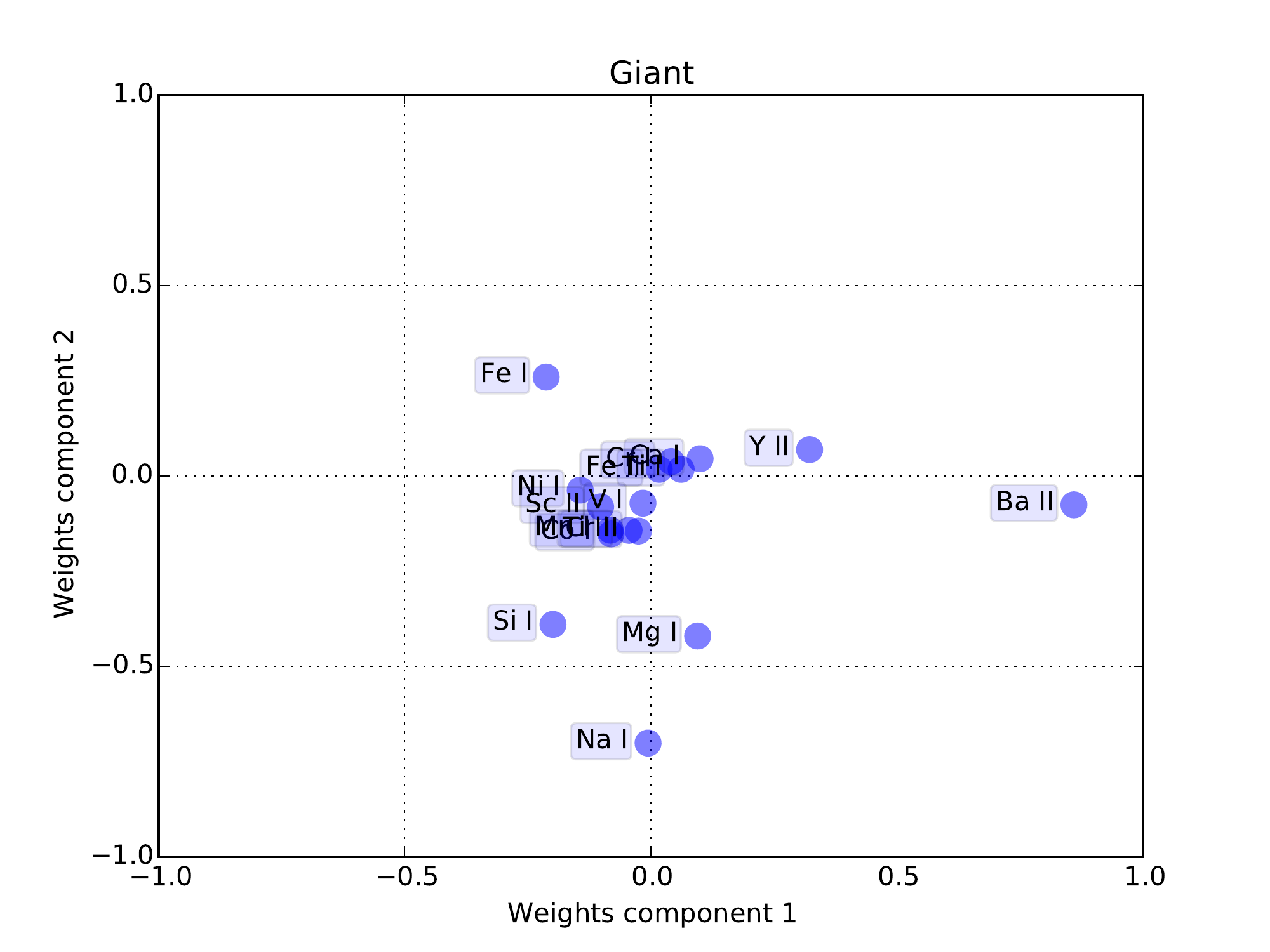}
    \end{centering}
    \caption{PCA weights of the elements per stellar subgroup.}
    \label{fig:elements}
\end{figure}

A principal component analysis can help us to understand the role of the measured abundances by looking at the weights assigned to each one. Elements with similar weights have similar behaviors and they do not contribute significantly to differentiate between clusters. In Fig \ref{fig:elements} we observe that the elements that contribute more to differentiate stars are the heavier elements (Y II, Ba II), Fe I, Mg I, Si I, and Na I. To fully take advantage of this information, it is desirable to include a higher number of stars in future studies, especially for the dwarf subgroup, to extend the analysis to include other elements and to correct NLTE effects.

\section{Conclusions}\label{s:conclusions}

We compiled 2,133 high-resolution spectra acquired with different instruments (i.e., NARVAL, HARPS, UVES) in the field of view of known open clusters. We implemented an automatic process based on iSpec to homogenize the observations, co-add them, derive atmospheric parameters and determine chemical abundances using a differential approach. After filtering low S/N spectra, non-members by radial velocity, non-FGK and/or chemically peculiar stars, we were left with a dataset of 177 stars covering 31 open clusters with abundances for 17 species corresponding to 14 different elements.

By slightly varying our continuum normalization process, we show how inhomogeneities in the spectral analysis imply systematic uncertainties in, for instance, the derived chemical abundances. Using the heterogeneous compilations from the literature to draw scientific conclusions about extensive topics such as the chemical history of our Galaxy is not recommended.

We identified distinct chemical signatures for stars in different evolutionary stages that belong to the same open cluster. The origin of these differences may be explained by NLTE effects (minimized for solar dwarfs thanks to the applied differential approach in the abundance determination), atomic diffusion, mixing processes and correlations from atmospheric parameter determinations. Regardless of whether these abundance enhancements are real or artificial, they have important implications for the chemical tagging technique when applied to stars in different evolutionary stages.

To evaluate the viability of the chemical tagging technique when analyzing a huge quantity of spectra in an automatized fashion, we performed experiments where we applied machine learning algorithms to blindly group stars based on their chemical abundances. We should note that our analysis was mainly limited to nearby clusters and it covers a narrow metallicity range. We found that the analyzed open clusters overlap in the chemical space for the 17 elemental abundances analyzed and it is not possible to completely recover co-natal stars (born from the same cloud at the same time). It is worth noting that chemical outliers were already discarded and the clustering analysis was performed individually in subgroups with stars in similar evolutionary stages. Thus, in a real scenario where the chemical tagging technique would be applied to a greater number of field stars, we expect to have a high level of overlapping that would severely affect the success rate of this technique for recovering co-natal aggregates.

In \cite{2014MNRAS.438.2753M}, the authors conducted the first blind chemical tagging experiment to find stellar groups from 714 field stars. They also found that the viability of finding co-natal groups was doubtful, but they claimed that the technique can still identify co-eval groups of stars (stars born at the same period). In our study, we observed a possible correlation between the first principal components and the stellar ages for giant stars. The door remains open for the possibility of using the chemical tagging technique to find co-eval aggregates.

It is intuitive to conclude that increases in chemical dimensionality lead to improvements in the clustering experiments, although the difficulty in deriving abundances for some elements (e.g., weak absorption lines, fewer lines, blended regions) at a given resolution could also yield greater uncertainties and potential scatter. We showed that not all the elements have the same discriminatory power \citep[as previous studies have forseen, such as][]{2002ARA&A..40..487F, 2013MNRAS.428.2321M}, some tend to act in concert while others contribute significantly such as the heavy n-capture element Ba. For future analysis, it would be interesting to include other elements such as La, Nd, and Eu, which are formed through similar processes that produce Ba (slow and rapid n-capture processes in low-mass AGB stars, \citealt{2001ApJ...557..802B}; and core-collapse supernovae, \citealt{2007ApJ...662...39K}). It would be also necessary to explore open clusters with lower metallicities, where less line blending could make different elements accessible.

There is also room to improve automatic analysis and spectral modeling, for instance, incorporating NLTE effects and averaged 3-D model atmosphere. Time dependent 3-D hydrodynamical models \citep{2013A&A...554A.118P} are still computationally too expensive as to use them for massive analysis, thus the averaged 3-D models are a good compromise. These improvements could reduce discrepancies among stars in different evolutionary stages and achieve a higher degree of success when recovering clusters using chemical abundances \citep{2012MNRAS.427...27B, 2012MNRAS.427...50L}.

\begin{acknowledgements}
    This work was partially supported by the Gaia Research for European Astronomy Training (GREAT-ITN) Marie Curie network, funded through the European Union Seventh Framework Programme [FP7/2007-2013] under grant agreement n. 264895.
    UH and AJK acknowledge support from the Swedish National Space Board (Rymdstyrelsen).
    ISR gratefully acknowledges the support provided by the Gemini-CONICYT project 32110029.
    All the software used in the data analysis were provided by the Open Source community.
\end{acknowledgements}


\bibliographystyle{bibtex/aa} 
\bibliography{References} 

\onllongtab{
\begin{landscape}

\footnotesize  
\tabcolsep=0.02cm
\setlength\LTleft{-30pt}            
\setlength\LTright{-30pt}           
\begin{longtable}{ll|cc|cc|cc|cc|cc|cc|cc|cc|cc|cc|cc|cc|cc|cc|cc|cc|cc|c}
\caption{Weighted mean abundances and dispersion per cluster and stellar subgroup.} \label{tab:abundances_startype}\\

\hline
\hline
&       &   \multicolumn{2}{c|}{[Ba II/Fe I]}         &   \multicolumn{2}{c|}{[Ca I/Fe I]}          &   \multicolumn{2}{c|}{[Co I/Fe I]}          &   \multicolumn{2}{c|}{[Cr I/Fe I]}          &   \multicolumn{2}{c|}{[Cr II/Fe I]}          &   \multicolumn{2}{c|}{[Fe I/H]}           &   \multicolumn{2}{c|}{[Fe II/Fe I]}          &   \multicolumn{2}{c|}{[Mg I/Fe I]}          &   \multicolumn{2}{c|}{[Mn I/Fe I]}          &   \multicolumn{2}{c|}{[Na I/Fe I]}          &   \multicolumn{2}{c|}{[Ni I/Fe I]}          &   \multicolumn{2}{c|}{[Sc II/Fe I]}          &   \multicolumn{2}{c|}{[Si I/Fe I]}          &   \multicolumn{2}{c|}{[Ti I/Fe I]}          &   \multicolumn{2}{c|}{[Ti II/Fe I]}          &   \multicolumn{2}{c|}{[V I/Fe I]}           &   \multicolumn{2}{c|}{[Y II/Fe I]}           &       \\

Cluster &   &   $\bar{x}_\mathrm{w}$    &   $\sigma_\mathrm{w}$ &   $\bar{x}_\mathrm{w}$    &   $\sigma_\mathrm{w}$ &   $\bar{x}_\mathrm{w}$    &   $\sigma_\mathrm{w}$ &   $\bar{x}_\mathrm{w}$    &   $\sigma_\mathrm{w}$ &   $\bar{x}_\mathrm{w}$    &   $\sigma_\mathrm{w}$ &   $\bar{x}_\mathrm{w}$    &   $\sigma_\mathrm{w}$ &   $\bar{x}_\mathrm{w}$    &   $\sigma_\mathrm{w}$ &   $\bar{x}_\mathrm{w}$    &   $\sigma_\mathrm{w}$ &   $\bar{x}_\mathrm{w}$    &   $\sigma_\mathrm{w}$ &   $\bar{x}_\mathrm{w}$    &   $\sigma_\mathrm{w}$ &   $\bar{x}_\mathrm{w}$    &   $\sigma_\mathrm{w}$ &   $\bar{x}_\mathrm{w}$    &   $\sigma_\mathrm{w}$ &   $\bar{x}_\mathrm{w}$    &   $\sigma_\mathrm{w}$ &   $\bar{x}_\mathrm{w}$    &   $\sigma_\mathrm{w}$ &   $\bar{x}_\mathrm{w}$    &   $\sigma_\mathrm{w}$ &   $\bar{x}_\mathrm{w}$    &   $\sigma_\mathrm{w}$ &   $\bar{x}_\mathrm{w}$    &   $\sigma_\mathrm{w}$ &   $\star$ \\
\hline
\endfirsthead
\caption{Continued.} \\
\hline
    &       &   \multicolumn{2}{c|}{Ba II}           &   \multicolumn{2}{c|}{Ca I}           &   \multicolumn{2}{c|}{Co I}           &   \multicolumn{2}{c|}{Cr I}           &   \multicolumn{2}{c|}{Cr II}           &   \multicolumn{2}{c|}{Fe I}           &   \multicolumn{2}{c|}{Fe II}           &   \multicolumn{2}{c|}{Mg I}           &   \multicolumn{2}{c|}{Mn I}           &   \multicolumn{2}{c|}{Na I}           &   \multicolumn{2}{c|}{Ni I}           &   \multicolumn{2}{c|}{Sc II}           &   \multicolumn{2}{c|}{Si I}           &   \multicolumn{2}{c|}{Ti I}           &   \multicolumn{2}{c|}{Ti II}           &   \multicolumn{2}{c|}{V I}            &   \multicolumn{2}{c|}{Y II}            &   \\
Cluster &       &   $\bar{x}_\mathrm{w}$    &   $\sigma_\mathrm{w}$ &   $\bar{x}_\mathrm{w}$    &   $\sigma_\mathrm{w}$ &   $\bar{x}_\mathrm{w}$    &   $\sigma_\mathrm{w}$ &   $\bar{x}_\mathrm{w}$    &   $\sigma_\mathrm{w}$ &   $\bar{x}_\mathrm{w}$    &   $\sigma_\mathrm{w}$ &   $\bar{x}_\mathrm{w}$    &   $\sigma_\mathrm{w}$ &   $\bar{x}_\mathrm{w}$    &   $\sigma_\mathrm{w}$ &   $\bar{x}_\mathrm{w}$    &   $\sigma_\mathrm{w}$ &   $\bar{x}_\mathrm{w}$    &   $\sigma_\mathrm{w}$ &   $\bar{x}_\mathrm{w}$    &   $\sigma_\mathrm{w}$ &   $\bar{x}_\mathrm{w}$    &   $\sigma_\mathrm{w}$ &   $\bar{x}_\mathrm{w}$    &   $\sigma_\mathrm{w}$ &   $\bar{x}_\mathrm{w}$    &   $\sigma_\mathrm{w}$ &   $\bar{x}_\mathrm{w}$    &   $\sigma_\mathrm{w}$ &   $\bar{x}_\mathrm{w}$    &   $\sigma_\mathrm{w}$ &   $\bar{x}_\mathrm{w}$    &   $\sigma_\mathrm{w}$ &   $\bar{x}_\mathrm{w}$    &   $\sigma_\mathrm{w}$ &   $\star$ \\

\hline
\endhead
\hline
\endfoot
\hline
\endlastfoot

        \hline

\footnotesize{Collinder350} &   \scriptsize{G}  &   0.29    &   0.17    &   0.06    &   0.11    &   -0.01   &   0.06    &   0.05    &   0.14    &   0.06    &   0.06    &   -0.10   &   0.10    &   0.00    &   0.12    &   0.08    &   0.03    &   0.02    &   0.07    &   0.25    &   0.05    &   -0.06   &   0.11    &   -0.03   &   0.07    &   0.10    &   0.06    &   0.03    &   0.14    &   0.02    &   0.11    &   0.08    &   0.07    &   0.09    &   0.17    &   1   \\
\footnotesize{IC2714}   &   \scriptsize{G}  &   0.20    &   0.06    &   0.05    &   0.02    &   0.00    &   0.01    &   0.02    &   0.02    &   0.03    &   0.03    &   -0.12   &   0.02    &   -0.01   &   0.03    &   0.11    &   0.03    &   0.04    &   0.01    &   0.28    &   0.02    &   -0.05   &   0.01    &   -0.02   &   0.02    &   0.12    &   0.02    &   0.03    &   0.01    &   0.01    &   0.01    &   0.08    &   0.02    &   0.07    &   0.02    &   6   \\
\footnotesize{IC4651}   &   \scriptsize{D}  &   -0.04   &   0.05    &   0.05    &   0.04    &   0.04    &   0.03    &   0.00    &   0.02    &   0.07    &   0.02    &   0.02    &   0.03    &   0.04    &   0.02    &   0.06    &   0.02    &   0.04    &   0.03    &   0.01    &   0.02    &   -0.01   &   0.02    &   -0.04   &   0.01    &   0.08    &   0.02    &   0.00    &   0.03    &   -0.04   &   0.01    &   0.06    &   0.04    &   0.01    &   0.03    &   6   \\
\footnotesize{IC4651}   &   \scriptsize{G}  &   -0.03   &   0.03    &   0.00    &   0.06    &   0.03    &   0.01    &   0.02    &   0.01    &   0.02    &   0.03    &   -0.02   &   0.04    &   -0.03   &   0.02    &   0.07    &   0.04    &   0.07    &   0.01    &   0.30    &   0.05    &   -0.01   &   0.03    &   0.04    &   0.00    &   0.17    &   0.04    &   0.02    &   0.00    &   0.03    &   0.02    &   0.07    &   0.00    &   0.00    &   0.01    &   2   \\
\footnotesize{IC4756}   &   \scriptsize{D}  &   0.16    &   0.07    &   0.05    &   0.05    &   -0.02   &   0.05    &   0.00    &   0.11    &   0.03    &   0.04    &   -0.02   &   0.06    &   0.04    &   0.08    &   0.04    &   0.03    &   -0.05   &   0.06    &   -0.12   &   0.04    &   -0.08   &   0.08    &   -0.05   &   0.05    &   0.02    &   0.03    &   0.00    &   0.09    &   -0.02   &   0.08    &   0.08    &   0.22    &   0.08    &   0.13    &   1   \\
\footnotesize{IC4756}   &   \scriptsize{G}  &   0.15    &   0.01    &   0.07    &   0.01    &   -0.01   &   0.00    &   0.02    &   0.01    &   0.03    &   0.01    &   -0.11   &   0.01    &   -0.01   &   0.01    &   0.09    &   0.01    &   0.02    &   0.00    &   0.23    &   0.03    &   -0.05   &   0.00    &   0.00    &   0.01    &   0.09    &   0.01    &   0.04    &   0.00    &   0.01    &   0.00    &   0.05    &   0.01    &   0.08    &   0.01    &   3   \\
\footnotesize{M67}  &   \scriptsize{D}  &   -0.13   &   0.06    &   0.04    &   0.04    &   0.06    &   0.02    &   -0.01   &   0.02    &   0.01    &   0.02    &   -0.04   &   0.04    &   0.00    &   0.03    &   0.07    &   0.02    &   0.02    &   0.03    &   0.01    &   0.02    &   -0.01   &   0.01    &   -0.01   &   0.02    &   0.06    &   0.02    &   0.02    &   0.02    &   -0.01   &   0.02    &   0.07    &   0.03    &   0.01    &   0.04    &   28  \\
\footnotesize{M67}  &   \scriptsize{G}  &   -0.16   &   0.03    &   0.02    &   0.01    &   0.03    &   0.01    &   0.00    &   0.01    &   0.03    &   0.02    &   -0.07   &   0.04    &   -0.01   &   0.02    &   0.14    &   0.05    &   0.05    &   0.01    &   0.21    &   0.05    &   0.00    &   0.01    &   0.04    &   0.01    &   0.16    &   0.04    &   0.02    &   0.01    &   0.03    &   0.02    &   0.07    &   0.01    &   -0.05   &   0.03    &   14  \\
\footnotesize{Melotte111}   &   \scriptsize{D}  &   0.11    &   0.03    &   0.04    &   0.01    &   -0.02   &   0.01    &   0.02    &   0.01    &   0.06    &   0.01    &   -0.07   &   0.02    &   -0.01   &   0.03    &   0.01    &   0.02    &   0.01    &   0.01    &   -0.05   &   0.02    &   -0.08   &   0.01    &   -0.06   &   0.02    &   0.04    &   0.02    &   0.02    &   0.02    &   -0.04   &   0.03    &   0.06    &   0.02    &   0.05    &   0.03    &   5   \\
\footnotesize{Melotte20}    &   \scriptsize{D}  &   0.07    &   0.13    &   0.05    &   0.07    &   0.05    &   0.09    &   0.01    &   0.14    &   0.06    &   0.06    &   0.03    &   0.11    &   0.05    &   0.11    &   0.05    &   0.06    &   -0.02   &   0.08    &   -0.09   &   0.06    &   -0.07   &   0.11    &   -0.07   &   0.04    &   0.05    &   0.05    &   -0.02   &   0.15    &   -0.08   &   0.10    &   0.14    &   0.24    &   0.08    &   0.16    &   1   \\
\footnotesize{Melotte22}    &   \scriptsize{D}  &   0.02    &   0.11    &   0.04    &   0.10    &   0.01    &   0.14    &   0.03    &   0.15    &   0.09    &   0.14    &   -0.06   &   0.12    &   0.00    &   0.11    &   0.06    &   0.05    &   0.01    &   0.10    &   -0.04   &   0.11    &   -0.08   &   0.12    &   -0.08   &   0.09    &   0.06    &   0.06    &   -0.02   &   0.18    &   -0.07   &   0.14    &   0.17    &   0.19    &   0.05    &   0.19    &   1   \\
\footnotesize{Melotte71}    &   \scriptsize{G}  &   0.25    &   0.04    &   0.02    &   0.15    &   -0.02   &   0.12    &   0.06    &   0.19    &   -0.09   &   0.10    &   -0.09   &   0.17    &   -0.02   &   0.20    &   0.12    &   0.16    &   0.00    &   0.11    &   0.10    &   0.15    &   -0.07   &   0.17    &   -0.04   &   0.10    &   0.02    &   0.10    &   0.04    &   0.17    &   0.01    &   0.20    &   -0.01   &   0.16    &   0.04    &   0.20    &   1   \\
\footnotesize{NGC1817}  &   \scriptsize{G}  &   0.19    &   0.02    &   0.08    &   0.03    &   -0.03   &   0.03    &   0.02    &   0.01    &   -0.04   &   0.01    &   -0.23   &   0.02    &   -0.06   &   0.01    &   0.08    &   0.01    &   -0.03   &   0.03    &   0.13    &   0.02    &   -0.06   &   0.01    &   0.00    &   0.01    &   0.08    &   0.02    &   0.05    &   0.02    &   -0.03   &   0.03    &   0.04    &   0.03    &   0.02    &   0.06    &   3   \\
\footnotesize{NGC2099}  &   \scriptsize{G}  &   0.05    &   0.13    &   0.04    &   0.25    &   -0.04   &   0.14    &   0.08    &   0.19    &   0.03    &   0.08    &   -0.03   &   0.20    &   -0.05   &   0.20    &   0.05    &   0.07    &   -0.03   &   0.16    &   0.14    &   0.19    &   -0.02   &   0.19    &   -0.07   &   0.12    &   0.08    &   0.18    &   -0.03   &   0.24    &   -0.01   &   0.17    &   -0.01   &   0.18    &   0.11    &   0.23    &   1   \\
\footnotesize{NGC2251}  &   \scriptsize{G}  &   0.21    &   0.04    &   0.04    &   0.11    &   -0.03   &   0.06    &   0.04    &   0.13    &   0.01    &   0.06    &   -0.09   &   0.11    &   -0.06   &   0.14    &   0.09    &   0.04    &   0.01    &   0.08    &   0.24    &   0.09    &   -0.06   &   0.12    &   -0.04   &   0.07    &   0.12    &   0.13    &   0.00    &   0.15    &   -0.01   &   0.13    &   0.02    &   0.09    &   0.08    &   0.19    &   1   \\
\footnotesize{NGC2360}  &   \scriptsize{G}  &   0.20    &   0.02    &   0.07    &   0.01    &   -0.02   &   0.01    &   0.02    &   0.01    &   -0.02   &   0.01    &   -0.11   &   0.03    &   -0.03   &   0.01    &   0.07    &   0.04    &   0.01    &   0.02    &   0.18    &   0.04    &   -0.06   &   0.00    &   0.00    &   0.01    &   0.08    &   0.02    &   0.04    &   0.00    &   0.02    &   0.01    &   0.04    &   0.01    &   0.07    &   0.01    &   7   \\
\footnotesize{NGC2423}  &   \scriptsize{G}  &   0.07    &   0.01    &   0.05    &   0.01    &   -0.02   &   0.00    &   0.02    &   0.00    &   0.03    &   0.01    &   0.02    &   0.02    &   -0.02   &   0.01    &   0.02    &   0.04    &   0.01    &   0.03    &   0.19    &   0.05    &   -0.04   &   0.00    &   0.01    &   0.01    &   0.11    &   0.00    &   0.01    &   0.00    &   0.02    &   0.01    &   0.03    &   0.00    &   0.05    &   0.01    &   2   \\
\footnotesize{NGC2447}  &   \scriptsize{D}  &   0.25    &   0.03    &   0.05    &   0.02    &   0.00    &   0.02    &   0.01    &   0.01    &   0.04    &   0.01    &   -0.12   &   0.01    &   0.00    &   0.02    &   0.02    &   0.01    &   -0.04   &   0.02    &   -0.09   &   0.01    &   -0.09   &   0.00    &   -0.07   &   0.02    &   0.03    &   0.01    &   0.01    &   0.02    &   -0.04   &   0.02    &   0.06    &   0.02    &   0.10    &   0.03    &   3   \\
\footnotesize{NGC2447}  &   \scriptsize{G}  &   0.20    &   0.04    &   0.08    &   0.01    &   -0.01   &   0.01    &   0.02    &   0.01    &   0.01    &   0.02    &   -0.13   &   0.02    &   -0.01   &   0.03    &   0.10    &   0.03    &   0.00    &   0.02    &   0.24    &   0.02    &   -0.05   &   0.01    &   0.00    &   0.01    &   0.10    &   0.02    &   0.05    &   0.01    &   0.02    &   0.02    &   0.04    &   0.01    &   0.10    &   0.02    &   3   \\
\footnotesize{NGC2477}  &   \scriptsize{G}  &   0.04    &   0.03    &   0.05    &   0.01    &   -0.01   &   0.01    &   0.03    &   0.02    &   0.01    &   0.03    &   -0.02   &   0.02    &   -0.02   &   0.03    &   0.04    &   0.02    &   0.04    &   0.02    &   0.21    &   0.02    &   -0.04   &   0.01    &   -0.02   &   0.01    &   0.11    &   0.02    &   0.01    &   0.01    &   0.01    &   0.02    &   0.07    &   0.01    &   0.04    &   0.02    &   29  \\
\footnotesize{NGC2539}  &   \scriptsize{G}  &   0.11    &   0.04    &   0.05    &   0.01    &   -0.02   &   0.01    &   0.02    &   0.01    &   0.01    &   0.02    &   -0.04   &   0.03    &   -0.02   &   0.02    &   0.04    &   0.03    &   0.04    &   0.02    &   0.24    &   0.02    &   -0.05   &   0.01    &   -0.01   &   0.02    &   0.11    &   0.02    &   0.01    &   0.01    &   0.01    &   0.02    &   0.04    &   0.02    &   0.07    &   0.01    &   7   \\
\footnotesize{NGC2547}  &   \scriptsize{D}  &   0.08    &   0.08    &   0.13    &   0.07    &   0.03    &   0.14    &   0.07    &   0.13    &   0.15    &   0.11    &   -0.13   &   0.14    &   -0.05   &   0.17    &   0.00    &   0.03    &   0.08    &   0.06    &   -0.01   &   0.10    &   -0.09   &   0.12    &   -0.08   &   0.12    &   0.07    &   0.05    &   0.04    &   0.18    &   -0.07   &   0.10    &   0.20    &   0.14    &   0.05    &   0.22    &   1   \\
\footnotesize{NGC2567}  &   \scriptsize{G}  &   0.23    &   0.05    &   0.07    &   0.02    &   0.00    &   0.01    &   0.02    &   0.01    &   0.01    &   0.02    &   -0.14   &   0.01    &   -0.01   &   0.03    &   0.10    &   0.03    &   0.02    &   0.02    &   0.24    &   0.02    &   -0.06   &   0.02    &   -0.02   &   0.02    &   0.11    &   0.02    &   0.04    &   0.02    &   0.03    &   0.01    &   0.07    &   0.03    &   0.10    &   0.02    &   4   \\
\footnotesize{NGC2632}  &   \scriptsize{D}  &   -0.09   &   0.04    &   0.03    &   0.03    &   0.02    &   0.03    &   0.03    &   0.02    &   0.05    &   0.03    &   0.11    &   0.02    &   0.05    &   0.01    &   -0.01   &   0.02    &   0.06    &   0.02    &   0.03    &   0.04    &   -0.02   &   0.01    &   -0.05   &   0.02    &   0.05    &   0.02    &   -0.02   &   0.02    &   -0.07   &   0.01    &   0.07    &   0.05    &   -0.01   &   0.03    &   11  \\
\footnotesize{NGC2632}  &   \scriptsize{G}  &   -0.01   &   0.01    &   0.04    &   0.01    &   0.01    &   0.01    &   0.04    &   0.01    &   0.04    &   0.01    &   0.07    &   0.01    &   0.00    &   0.01    &   0.04    &   0.05    &   0.04    &   0.02    &   0.34    &   0.01    &   0.01    &   0.00    &   0.04    &   0.00    &   0.18    &   0.01    &   -0.02   &   0.00    &   -0.01   &   0.04    &   0.04    &   0.01    &   0.02    &   0.03    &   2   \\
\footnotesize{NGC3532}  &   \scriptsize{G}  &   0.21    &   0.02    &   0.05    &   0.02    &   0.00    &   0.01    &   0.04    &   0.01    &   0.04    &   0.05    &   -0.11   &   0.04    &   -0.03   &   0.02    &   0.08    &   0.04    &   0.03    &   0.02    &   0.27    &   0.05    &   -0.05   &   0.01    &   0.00    &   0.02    &   0.12    &   0.03    &   0.04    &   0.02    &   0.02    &   0.02    &   0.09    &   0.05    &   0.08    &   0.02    &   3   \\
\footnotesize{NGC3680}  &   \scriptsize{D}  &   0.00    &   0.09    &   0.07    &   0.00    &   0.01    &   0.00    &   0.01    &   0.04    &   0.06    &   0.01    &   -0.10   &   0.04    &   0.01    &   0.00    &   0.07    &   0.04    &   -0.02   &   0.04    &   -0.03   &   0.01    &   -0.06   &   0.01    &   -0.04   &   0.01    &   0.05    &   0.02    &   0.01    &   0.04    &   -0.02   &   0.01    &   0.09    &   0.07    &   0.07    &   0.01    &   2   \\
\footnotesize{NGC3680}  &   \scriptsize{G}  &   0.08    &   0.04    &   0.02    &   0.01    &   -0.01   &   0.02    &   0.00    &   0.01    &   0.02    &   0.03    &   -0.13   &   0.04    &   -0.01   &   0.03    &   0.14    &   0.05    &   0.00    &   0.04    &   0.15    &   0.03    &   -0.04   &   0.01    &   0.04    &   0.03    &   0.17    &   0.02    &   0.02    &   0.01    &   0.04    &   0.03    &   0.04    &   0.01    &   0.03    &   0.02    &   3   \\
\footnotesize{NGC4349}  &   \scriptsize{G}  &   0.15    &   0.03    &   0.05    &   0.01    &   -0.01   &   0.01    &   0.01    &   0.01    &   0.01    &   0.02    &   -0.16   &   0.03    &   -0.03   &   0.02    &   0.12    &   0.02    &   0.03    &   0.02    &   0.23    &   0.01    &   -0.05   &   0.01    &   -0.03   &   0.01    &   0.11    &   0.02    &   0.02    &   0.01    &   -0.02   &   0.01    &   0.06    &   0.02    &   0.09    &   0.01    &   3   \\
\footnotesize{NGC5822}  &   \scriptsize{G}  &   0.28    &   0.00    &   0.07    &   0.02    &   0.00    &   0.02    &   0.03    &   0.01    &   0.07    &   0.04    &   -0.12   &   0.02    &   0.02    &   0.05    &   0.08    &   0.00    &   0.03    &   0.00    &   0.21    &   0.02    &   -0.05   &   0.00    &   0.00    &   0.02    &   0.10    &   0.03    &   0.04    &   0.02    &   0.02    &   0.03    &   0.07    &   0.03    &   0.16    &   0.08    &   2   \\
\footnotesize{NGC6475}  &   \scriptsize{D}  &   0.12    &   0.09    &   0.01    &   0.05    &   -0.02   &   0.04    &   0.03    &   0.08    &   0.05    &   0.03    &   0.04    &   0.05    &   0.01    &   0.08    &   -0.03   &   0.02    &   0.03    &   0.03    &   -0.12   &   0.06    &   -0.09   &   0.06    &   -0.10   &   0.06    &   0.00    &   0.04    &   0.00    &   0.10    &   -0.08   &   0.07    &   0.04    &   0.05    &   0.05    &   0.09    &   1   \\
\footnotesize{NGC6494}  &   \scriptsize{G}  &   0.17    &   0.03    &   0.04    &   0.01    &   0.01    &   0.01    &   0.02    &   0.01    &   0.02    &   0.02    &   -0.12   &   0.02    &   -0.02   &   0.02    &   0.14    &   0.02    &   0.05    &   0.01    &   0.35    &   0.03    &   -0.05   &   0.00    &   0.00    &   0.02    &   0.18    &   0.02    &   0.01    &   0.00    &   0.02    &   0.02    &   0.07    &   0.03    &   0.06    &   0.01    &   4   \\
\footnotesize{NGC6633}  &   \scriptsize{G}  &   0.20    &   0.02    &   0.06    &   0.01    &   -0.01   &   0.00    &   0.02    &   0.00    &   0.01    &   0.01    &   -0.08   &   0.02    &   -0.01   &   0.01    &   0.09    &   0.04    &   0.03    &   0.00    &   0.23    &   0.04    &   -0.06   &   0.00    &   -0.01   &   0.01    &   0.08    &   0.02    &   0.03    &   0.01    &   0.00    &   0.02    &   0.05    &   0.01    &   0.08    &   0.01    &   4   \\
\footnotesize{NGC6705}  &   \scriptsize{G}  &   0.01    &   0.05    &   0.03    &   0.01    &   0.04    &   0.01    &   0.03    &   0.01    &   0.07    &   0.04    &   -0.01   &   0.02    &   -0.01   &   0.01    &   0.14    &   0.05    &   0.06    &   0.02    &   0.41    &   0.00    &   -0.02   &   0.01    &   0.02    &   0.04    &   0.24    &   0.02    &   -0.01   &   0.00    &   0.08    &   0.01    &   0.06    &   0.00    &   -0.02   &   0.00    &   2   \\
\footnotesize{NGC6811}  &   \scriptsize{G}  &   0.26    &   0.01    &   0.05    &   0.00    &   0.00    &   0.00    &   0.04    &   0.01    &   0.04    &   0.00    &   -0.10   &   0.01    &   0.04    &   0.02    &   0.06    &   0.01    &   0.02    &   0.01    &   0.16    &   0.00    &   -0.07   &   0.00    &   -0.01   &   0.00    &   0.06    &   0.02    &   0.05    &   0.01    &   -0.01   &   0.01    &   0.08    &   0.00    &   0.12    &   0.01    &   2   \\
\footnotesize{NGC6940}  &   \scriptsize{G}  &   0.04    &   0.07    &   0.03    &   0.09    &   -0.04   &   0.06    &   0.01    &   0.09    &   0.00    &   0.07    &   0.04    &   0.09    &   -0.05   &   0.11    &   0.02    &   0.02    &   0.02    &   0.08    &   0.25    &   0.08    &   -0.05   &   0.10    &   -0.02   &   0.04    &   0.13    &   0.13    &   -0.01   &   0.13    &   -0.02   &   0.11    &   0.03    &   0.08    &   0.04    &   0.17    &   1   \\
\footnotesize{NGC752}   &   \scriptsize{G}  &   0.10    &   0.01    &   0.04    &   0.01    &   -0.04   &   0.01    &   0.01    &   0.00    &   0.00    &   0.01    &   -0.07   &   0.01    &   -0.04   &   0.01    &   0.05    &   0.01    &   0.00    &   0.01    &   0.15    &   0.01    &   -0.06   &   0.00    &   -0.01   &   0.02    &   0.10    &   0.01    &   0.01    &   0.01    &   0.01    &   0.01    &   0.03    &   0.01    &   0.04    &   0.01    &   7   \\

\end{longtable}
    \tablefoot{Subgroups correspond to dwarfs (D) and giants (G). The number of stars is indicated in the last column ($\star$). }
\end{landscape}
}

\onllongtab{
\begin{landscape}

\footnotesize  
\tabcolsep=0.02cm
\setlength\LTleft{-30pt}            
\setlength\LTright{-30pt}           
\begin{longtable}{l|cc|cc|cc|cc|cc|cc|cc|cc|cc|cc|cc|cc|cc|cc|cc|cc|cc|c}
\caption{Weighted mean abundances and dispersion per cluster.} \label{tab:abundances_cluster} \\

\hline
\hline

    &   \multicolumn{2}{c|}{[Ba II/Fe I]}          &   \multicolumn{2}{c|}{[Ca I/Fe I]}          &   \multicolumn{2}{c|}{[Co I/Fe I]}          &   \multicolumn{2}{c|}{[Cr I/Fe I]}          &   \multicolumn{2}{c|}{[Cr II/Fe I]}          &   \multicolumn{2}{c|}{[Fe I/H]}           &   \multicolumn{2}{c|}{[Fe II/Fe I]}          &   \multicolumn{2}{c|}{[Mg I/Fe I]}          &   \multicolumn{2}{c|}{[Mn I/Fe I]}          &   \multicolumn{2}{c|}{[Na I/Fe I]}          &   \multicolumn{2}{c|}{[Ni I/Fe I]}          &   \multicolumn{2}{c|}{[Sc II/Fe I]}          &   \multicolumn{2}{c|}{[Si I/Fe I]}          &   \multicolumn{2}{c|}{[Ti I/Fe I]}          &   \multicolumn{2}{c|}{[Ti II/Fe I]}          &   \multicolumn{2}{c|}{[V I/Fe I]}           &   \multicolumn{2}{c|}{[Y II/Fe I]}           &       \\
Cluster &   $\bar{x}_\mathrm{w}$    &   $\sigma_\mathrm{w}$ &   $\bar{x}_\mathrm{w}$    &   $\sigma_\mathrm{w}$ &   $\bar{x}_\mathrm{w}$    &   $\sigma_\mathrm{w}$ &   $\bar{x}_\mathrm{w}$    &   $\sigma_\mathrm{w}$ &   $\bar{x}_\mathrm{w}$    &   $\sigma_\mathrm{w}$ &   $\bar{x}_\mathrm{w}$    &   $\sigma_\mathrm{w}$ &   $\bar{x}_\mathrm{w}$    &   $\sigma_\mathrm{w}$ &   $\bar{x}_\mathrm{w}$    &   $\sigma_\mathrm{w}$ &   $\bar{x}_\mathrm{w}$    &   $\sigma_\mathrm{w}$ &   $\bar{x}_\mathrm{w}$    &   $\sigma_\mathrm{w}$ &   $\bar{x}_\mathrm{w}$    &   $\sigma_\mathrm{w}$ &   $\bar{x}_\mathrm{w}$    &   $\sigma_\mathrm{w}$ &   $\bar{x}_\mathrm{w}$    &   $\sigma_\mathrm{w}$ &   $\bar{x}_\mathrm{w}$    &   $\sigma_\mathrm{w}$ &   $\bar{x}_\mathrm{w}$    &   $\sigma_\mathrm{w}$ &   $\bar{x}_\mathrm{w}$    &   $\sigma_\mathrm{w}$ &   $\bar{x}_\mathrm{w}$    &   $\sigma_\mathrm{w}$ &   $\star$ \\

\hline
\endfirsthead
\caption{Continued.} \\
\hline

    &   \multicolumn{2}{c|}{[Ba II/Fe I]}          &   \multicolumn{2}{c|}{[Ca I/Fe I]}          &   \multicolumn{2}{c|}{[Co I/Fe I]}          &   \multicolumn{2}{c|}{[Cr I/Fe I]}          &   \multicolumn{2}{c|}{[Cr II/Fe I]}          &   \multicolumn{2}{c|}{[Fe I/H]}           &   \multicolumn{2}{c|}{[Fe II/Fe I]}          &   \multicolumn{2}{c|}{[Mg I/Fe I]}          &   \multicolumn{2}{c|}{[Mn I/Fe I]}          &   \multicolumn{2}{c|}{[Na I/Fe I]}          &   \multicolumn{2}{c|}{[Ni I/Fe I]}          &   \multicolumn{2}{c|}{[Sc II/Fe I]}          &   \multicolumn{2}{c|}{[Si I/Fe I]}          &   \multicolumn{2}{c|}{[Ti I/Fe I]}          &   \multicolumn{2}{c|}{[Ti II/Fe I]}          &   \multicolumn{2}{c|}{[V I/Fe I]}           &   \multicolumn{2}{c|}{[Y II/Fe I]}           &       \\
Cluster &   $\bar{x}_\mathrm{w}$    &   $\sigma_\mathrm{w}$ &   $\bar{x}_\mathrm{w}$    &   $\sigma_\mathrm{w}$ &   $\bar{x}_\mathrm{w}$    &   $\sigma_\mathrm{w}$ &   $\bar{x}_\mathrm{w}$    &   $\sigma_\mathrm{w}$ &   $\bar{x}_\mathrm{w}$    &   $\sigma_\mathrm{w}$ &   $\bar{x}_\mathrm{w}$    &   $\sigma_\mathrm{w}$ &   $\bar{x}_\mathrm{w}$    &   $\sigma_\mathrm{w}$ &   $\bar{x}_\mathrm{w}$    &   $\sigma_\mathrm{w}$ &   $\bar{x}_\mathrm{w}$    &   $\sigma_\mathrm{w}$ &   $\bar{x}_\mathrm{w}$    &   $\sigma_\mathrm{w}$ &   $\bar{x}_\mathrm{w}$    &   $\sigma_\mathrm{w}$ &   $\bar{x}_\mathrm{w}$    &   $\sigma_\mathrm{w}$ &   $\bar{x}_\mathrm{w}$    &   $\sigma_\mathrm{w}$ &   $\bar{x}_\mathrm{w}$    &   $\sigma_\mathrm{w}$ &   $\bar{x}_\mathrm{w}$    &   $\sigma_\mathrm{w}$ &   $\bar{x}_\mathrm{w}$    &   $\sigma_\mathrm{w}$ &   $\bar{x}_\mathrm{w}$    &   $\sigma_\mathrm{w}$ &   $\star$ \\

\hline
\endhead
\hline
\endfoot
\hline
\endlastfoot

        \hline

\footnotesize{Collinder350} &   0.29    &   0.17    &   0.06    &   0.11    &   -0.01   &   0.06    &   0.05    &   0.14    &   0.06    &   0.06    &   -0.10   &   0.10    &   0.00    &   0.12    &   0.08    &   0.03    &   0.02    &   0.07    &   0.25    &   0.05    &   -0.06   &   0.11    &   -0.03   &   0.07    &   0.10    &   0.06    &   0.03    &   0.14    &   0.02    &   0.11    &   0.08    &   0.07    &   0.09    &   0.17    &   1   \\
\footnotesize{IC2714}   &   0.20    &   0.06    &   0.05    &   0.02    &   0.00    &   0.01    &   0.02    &   0.02    &   0.03    &   0.03    &   -0.12   &   0.02    &   -0.01   &   0.03    &   0.11    &   0.03    &   0.04    &   0.01    &   0.28    &   0.02    &   -0.05   &   0.01    &   -0.02   &   0.02    &   0.12    &   0.02    &   0.03    &   0.01    &   0.01    &   0.01    &   0.08    &   0.02    &   0.07    &   0.02    &   6   \\
\footnotesize{IC4651}   &   -0.03   &   0.03    &   0.04    &   0.04    &   0.03    &   0.02    &   0.01    &   0.02    &   0.06    &   0.03    &   0.01    &   0.03    &   0.02    &   0.03    &   0.06    &   0.02    &   0.05    &   0.03    &   0.09    &   0.14    &   -0.01   &   0.02    &   -0.01   &   0.04    &   0.08    &   0.02    &   0.00    &   0.03    &   -0.03   &   0.03    &   0.06    &   0.03    &   0.00    &   0.03    &   8   \\
\footnotesize{IC4756}   &   0.15    &   0.01    &   0.06    &   0.01    &   -0.01   &   0.00    &   0.02    &   0.01    &   0.03    &   0.01    &   -0.07   &   0.05    &   0.01    &   0.03    &   0.06    &   0.03    &   0.01    &   0.03    &   0.11    &   0.19    &   -0.06   &   0.02    &   -0.01   &   0.03    &   0.07    &   0.04    &   0.03    &   0.02    &   0.00    &   0.02    &   0.05    &   0.01    &   0.08    &   0.00    &   4   \\
\footnotesize{M67}  &   -0.16   &   0.03    &   0.03    &   0.03    &   0.05    &   0.02    &   -0.01   &   0.02    &   0.02    &   0.02    &   -0.05   &   0.04    &   0.00    &   0.03    &   0.11    &   0.05    &   0.03    &   0.03    &   0.03    &   0.05    &   0.00    &   0.01    &   0.02    &   0.03    &   0.06    &   0.03    &   0.02    &   0.02    &   0.00    &   0.03    &   0.07    &   0.02    &   -0.01   &   0.05    &   42  \\
\footnotesize{Melotte111}   &   0.11    &   0.03    &   0.04    &   0.01    &   -0.02   &   0.01    &   0.02    &   0.01    &   0.06    &   0.01    &   -0.07   &   0.02    &   -0.01   &   0.03    &   0.01    &   0.02    &   0.01    &   0.01    &   -0.05   &   0.02    &   -0.08   &   0.01    &   -0.06   &   0.02    &   0.04    &   0.02    &   0.02    &   0.02    &   -0.04   &   0.03    &   0.06    &   0.02    &   0.05    &   0.03    &   5   \\
\footnotesize{Melotte20}    &   0.07    &   0.13    &   0.05    &   0.07    &   0.05    &   0.09    &   0.01    &   0.14    &   0.06    &   0.06    &   0.03    &   0.11    &   0.05    &   0.11    &   0.05    &   0.06    &   -0.02   &   0.08    &   -0.09   &   0.06    &   -0.07   &   0.11    &   -0.07   &   0.04    &   0.05    &   0.05    &   -0.02   &   0.15    &   -0.08   &   0.10    &   0.14    &   0.24    &   0.08    &   0.16    &   1   \\
\footnotesize{Melotte22}    &   0.02    &   0.11    &   0.04    &   0.10    &   0.01    &   0.14    &   0.03    &   0.15    &   0.09    &   0.14    &   -0.06   &   0.12    &   0.00    &   0.11    &   0.06    &   0.05    &   0.01    &   0.10    &   -0.04   &   0.11    &   -0.08   &   0.12    &   -0.08   &   0.09    &   0.06    &   0.06    &   -0.02   &   0.18    &   -0.07   &   0.14    &   0.17    &   0.19    &   0.05    &   0.19    &   1   \\
\footnotesize{Melotte71}    &   0.25    &   0.04    &   0.02    &   0.15    &   -0.02   &   0.12    &   0.06    &   0.19    &   -0.09   &   0.10    &   -0.09   &   0.17    &   -0.02   &   0.20    &   0.12    &   0.16    &   0.00    &   0.11    &   0.10    &   0.15    &   -0.07   &   0.17    &   -0.04   &   0.10    &   0.02    &   0.10    &   0.04    &   0.17    &   0.01    &   0.20    &   -0.01   &   0.16    &   0.04    &   0.20    &   1   \\
\footnotesize{NGC1817}  &   0.19    &   0.02    &   0.08    &   0.03    &   -0.03   &   0.03    &   0.02    &   0.01    &   -0.04   &   0.01    &   -0.23   &   0.02    &   -0.06   &   0.01    &   0.08    &   0.01    &   -0.03   &   0.03    &   0.13    &   0.02    &   -0.06   &   0.01    &   0.00    &   0.01    &   0.08    &   0.02    &   0.05    &   0.02    &   -0.03   &   0.03    &   0.04    &   0.03    &   0.02    &   0.06    &   3   \\
\footnotesize{NGC2099}  &   0.05    &   0.13    &   0.04    &   0.25    &   -0.04   &   0.14    &   0.08    &   0.19    &   0.03    &   0.08    &   -0.03   &   0.20    &   -0.05   &   0.20    &   0.05    &   0.07    &   -0.03   &   0.16    &   0.14    &   0.19    &   -0.02   &   0.19    &   -0.07   &   0.12    &   0.08    &   0.18    &   -0.03   &   0.24    &   -0.01   &   0.17    &   -0.01   &   0.18    &   0.11    &   0.23    &   1   \\
\footnotesize{NGC2251}  &   0.21    &   0.04    &   0.04    &   0.11    &   -0.03   &   0.06    &   0.04    &   0.13    &   0.01    &   0.06    &   -0.09   &   0.11    &   -0.06   &   0.14    &   0.09    &   0.04    &   0.01    &   0.08    &   0.24    &   0.09    &   -0.06   &   0.12    &   -0.04   &   0.07    &   0.12    &   0.13    &   0.00    &   0.15    &   -0.01   &   0.13    &   0.02    &   0.09    &   0.08    &   0.19    &   1   \\
\footnotesize{NGC2360}  &   0.20    &   0.02    &   0.07    &   0.01    &   -0.02   &   0.01    &   0.02    &   0.01    &   -0.02   &   0.01    &   -0.11   &   0.03    &   -0.03   &   0.01    &   0.07    &   0.04    &   0.01    &   0.02    &   0.18    &   0.04    &   -0.06   &   0.00    &   0.00    &   0.01    &   0.08    &   0.02    &   0.04    &   0.00    &   0.02    &   0.01    &   0.04    &   0.01    &   0.07    &   0.01    &   7   \\
\footnotesize{NGC2423}  &   0.07    &   0.01    &   0.05    &   0.01    &   -0.02   &   0.00    &   0.02    &   0.00    &   0.03    &   0.01    &   0.02    &   0.02    &   -0.02   &   0.01    &   0.02    &   0.04    &   0.01    &   0.03    &   0.19    &   0.05    &   -0.04   &   0.00    &   0.01    &   0.01    &   0.11    &   0.00    &   0.01    &   0.00    &   0.02    &   0.01    &   0.03    &   0.00    &   0.05    &   0.01    &   2   \\
\footnotesize{NGC2447}  &   0.21    &   0.04    &   0.06    &   0.02    &   -0.01   &   0.01    &   0.01    &   0.01    &   0.02    &   0.03    &   -0.13   &   0.02    &   0.00    &   0.02    &   0.05    &   0.04    &   -0.02   &   0.03    &   0.11    &   0.17    &   -0.07   &   0.02    &   -0.01   &   0.03    &   0.07    &   0.04    &   0.03    &   0.02    &   -0.01   &   0.04    &   0.05    &   0.02    &   0.10    &   0.02    &   6   \\
\footnotesize{NGC2477}  &   0.04    &   0.03    &   0.05    &   0.01    &   -0.01   &   0.01    &   0.03    &   0.02    &   0.01    &   0.03    &   -0.02   &   0.02    &   -0.02   &   0.03    &   0.04    &   0.02    &   0.04    &   0.02    &   0.21    &   0.02    &   -0.04   &   0.01    &   -0.02   &   0.01    &   0.11    &   0.02    &   0.01    &   0.01    &   0.01    &   0.02    &   0.07    &   0.01    &   0.04    &   0.02    &   29  \\
\footnotesize{NGC2539}  &   0.11    &   0.04    &   0.05    &   0.01    &   -0.02   &   0.01    &   0.02    &   0.01    &   0.01    &   0.02    &   -0.04   &   0.03    &   -0.02   &   0.02    &   0.04    &   0.03    &   0.04    &   0.02    &   0.24    &   0.02    &   -0.05   &   0.01    &   -0.01   &   0.02    &   0.11    &   0.02    &   0.01    &   0.01    &   0.01    &   0.02    &   0.04    &   0.02    &   0.07    &   0.01    &   7   \\
\footnotesize{NGC2547}  &   0.08    &   0.08    &   0.13    &   0.07    &   0.03    &   0.14    &   0.07    &   0.13    &   0.15    &   0.11    &   -0.13   &   0.14    &   -0.05   &   0.17    &   0.00    &   0.03    &   0.08    &   0.06    &   -0.01   &   0.10    &   -0.09   &   0.12    &   -0.08   &   0.12    &   0.07    &   0.05    &   0.04    &   0.18    &   -0.07   &   0.10    &   0.20    &   0.14    &   0.05    &   0.22    &   1   \\
\footnotesize{NGC2567}  &   0.23    &   0.05    &   0.07    &   0.02    &   0.00    &   0.01    &   0.02    &   0.01    &   0.01    &   0.02    &   -0.14   &   0.01    &   -0.01   &   0.03    &   0.10    &   0.03    &   0.02    &   0.02    &   0.24    &   0.02    &   -0.06   &   0.02    &   -0.02   &   0.02    &   0.11    &   0.02    &   0.04    &   0.02    &   0.03    &   0.01    &   0.07    &   0.03    &   0.10    &   0.02    &   4   \\
\footnotesize{NGC2632}  &   -0.01   &   0.06    &   0.03    &   0.02    &   0.02    &   0.02    &   0.03    &   0.02    &   0.05    &   0.03    &   0.10    &   0.03    &   0.05    &   0.02    &   0.00    &   0.02    &   0.06    &   0.02    &   0.05    &   0.09    &   -0.02   &   0.02    &   -0.04   &   0.04    &   0.05    &   0.02    &   -0.02   &   0.02    &   -0.06   &   0.02    &   0.06    &   0.05    &   0.00    &   0.03    &   13  \\
\footnotesize{NGC3532}  &   0.21    &   0.02    &   0.05    &   0.02    &   0.00    &   0.01    &   0.04    &   0.01    &   0.04    &   0.05    &   -0.11   &   0.04    &   -0.03   &   0.02    &   0.08    &   0.04    &   0.03    &   0.02    &   0.27    &   0.05    &   -0.05   &   0.01    &   0.00    &   0.02    &   0.12    &   0.03    &   0.04    &   0.02    &   0.02    &   0.02    &   0.09    &   0.05    &   0.08    &   0.02    &   3   \\
\footnotesize{NGC3680}  &   0.06    &   0.06    &   0.04    &   0.03    &   -0.01   &   0.02    &   0.00    &   0.02    &   0.03    &   0.03    &   -0.11   &   0.04    &   0.00    &   0.02    &   0.14    &   0.05    &   0.00    &   0.03    &   0.05    &   0.10    &   -0.05   &   0.02    &   0.02    &   0.04    &   0.06    &   0.04    &   0.02    &   0.02    &   0.02    &   0.04    &   0.05    &   0.04    &   0.04    &   0.03    &   5   \\
\footnotesize{NGC4349}  &   0.15    &   0.03    &   0.05    &   0.01    &   -0.01   &   0.01    &   0.01    &   0.01    &   0.01    &   0.02    &   -0.16   &   0.03    &   -0.03   &   0.02    &   0.12    &   0.02    &   0.03    &   0.02    &   0.23    &   0.01    &   -0.05   &   0.01    &   -0.03   &   0.01    &   0.11    &   0.02    &   0.02    &   0.01    &   -0.02   &   0.01    &   0.06    &   0.02    &   0.09    &   0.01    &   3   \\
\footnotesize{NGC5822}  &   0.28    &   0.00    &   0.07    &   0.02    &   0.00    &   0.02    &   0.03    &   0.01    &   0.07    &   0.04    &   -0.12   &   0.02    &   0.02    &   0.05    &   0.08    &   0.00    &   0.03    &   0.00    &   0.21    &   0.02    &   -0.05   &   0.00    &   0.00    &   0.02    &   0.10    &   0.03    &   0.04    &   0.02    &   0.02    &   0.03    &   0.07    &   0.03    &   0.16    &   0.08    &   2   \\
\footnotesize{NGC6475}  &   0.12    &   0.09    &   0.01    &   0.05    &   -0.02   &   0.04    &   0.03    &   0.08    &   0.05    &   0.03    &   0.04    &   0.05    &   0.01    &   0.08    &   -0.03   &   0.02    &   0.03    &   0.03    &   -0.12   &   0.06    &   -0.09   &   0.06    &   -0.10   &   0.06    &   0.00    &   0.04    &   0.00    &   0.10    &   -0.08   &   0.07    &   0.04    &   0.05    &   0.05    &   0.09    &   1   \\
\footnotesize{NGC6494}  &   0.17    &   0.03    &   0.04    &   0.01    &   0.01    &   0.01    &   0.02    &   0.01    &   0.02    &   0.02    &   -0.12   &   0.02    &   -0.02   &   0.02    &   0.14    &   0.02    &   0.05    &   0.01    &   0.35    &   0.03    &   -0.05   &   0.00    &   0.00    &   0.02    &   0.18    &   0.02    &   0.01    &   0.00    &   0.02    &   0.02    &   0.07    &   0.03    &   0.06    &   0.01    &   4   \\
\footnotesize{NGC6633}  &   0.20    &   0.02    &   0.06    &   0.01    &   -0.01   &   0.00    &   0.02    &   0.00    &   0.01    &   0.01    &   -0.08   &   0.02    &   -0.01   &   0.01    &   0.09    &   0.04    &   0.03    &   0.00    &   0.23    &   0.04    &   -0.06   &   0.00    &   -0.01   &   0.01    &   0.08    &   0.02    &   0.03    &   0.01    &   0.00    &   0.02    &   0.05    &   0.01    &   0.08    &   0.01    &   4   \\
\footnotesize{NGC6705}  &   0.01    &   0.05    &   0.03    &   0.01    &   0.04    &   0.01    &   0.03    &   0.01    &   0.07    &   0.04    &   -0.01   &   0.02    &   -0.01   &   0.01    &   0.14    &   0.05    &   0.06    &   0.02    &   0.41    &   0.00    &   -0.02   &   0.01    &   0.02    &   0.04    &   0.24    &   0.02    &   -0.01   &   0.00    &   0.08    &   0.01    &   0.06    &   0.00    &   -0.02   &   0.00    &   2   \\
\footnotesize{NGC6811}  &   0.26    &   0.01    &   0.05    &   0.00    &   0.00    &   0.00    &   0.04    &   0.01    &   0.04    &   0.00    &   -0.10   &   0.01    &   0.04    &   0.02    &   0.06    &   0.01    &   0.02    &   0.01    &   0.16    &   0.00    &   -0.07   &   0.00    &   -0.01   &   0.00    &   0.06    &   0.02    &   0.05    &   0.01    &   -0.01   &   0.01    &   0.08    &   0.00    &   0.12    &   0.01    &   2   \\
\footnotesize{NGC6940}  &   0.04    &   0.07    &   0.03    &   0.09    &   -0.04   &   0.06    &   0.01    &   0.09    &   0.00    &   0.07    &   0.04    &   0.09    &   -0.05   &   0.11    &   0.02    &   0.02    &   0.02    &   0.08    &   0.25    &   0.08    &   -0.05   &   0.10    &   -0.02   &   0.04    &   0.13    &   0.13    &   -0.01   &   0.13    &   -0.02   &   0.11    &   0.03    &   0.08    &   0.04    &   0.17    &   1   \\
\footnotesize{NGC752}   &   0.10    &   0.01    &   0.04    &   0.01    &   -0.04   &   0.01    &   0.01    &   0.00    &   0.00    &   0.01    &   -0.07   &   0.01    &   -0.04   &   0.01    &   0.05    &   0.01    &   0.00    &   0.01    &   0.15    &   0.01    &   -0.06   &   0.00    &   -0.01   &   0.02    &   0.10    &   0.01    &   0.01    &   0.01    &   0.01    &   0.01    &   0.03    &   0.01    &   0.04    &   0.01    &   7   \\

\end{longtable}
    \tablefoot{ The number of stars is indicated in the last column ($\star$). }
\end{landscape}
}

\end{document}